\documentclass[12pt]{article}
\usepackage{graphicx}
\usepackage{amssymb,amsmath,epsfig}

\renewcommand{\theequation}{\arabic{section}.\arabic{equation}}

\begin{document}

\title{\bf Hot Plasma Modes Across Reissner-N{\"o}rdstrom de Sitter Horizon in a Veselago Medium}
\author{Ifra Noureen \thanks{ifra.noureen@gmail.com}\\
University of Management and Technology, Lahore, Pakistan.\\
H. Rizwana Kausar\thanks{rizwa\_math@yahoo.com}\\
Director, Centre for Applicable Mathematics \& Statistics,\\
University of Central Punjab, Lahore, Pakistan.}
\date{}
\maketitle
\begin{abstract}

In this manuscript, wave attributes of hot plasma
around Reissner-N{\"o}rdstrom-de Sitter (RN-dS) metric in a Veselago
medium are investigated. General relativistic magnetohydrodynamical
(GRMHD) equations in planar coordinates for the RN-dS horizon are
reformulated by implementation of ADM $3+1$ technique.
Further, perturbation scheme is used to arrive at linearly perturbed
GRMHD equations whose component form is used to attain dispersion
relations for rotating (non-magnetized and magnetized) plasma. Wave
propagation in hot plasma is explained by three
dimensional graphical representation of the wave number, refractive
index, its change with respect to angular velocity, phase and group
velocities. Finally, comparison of wave properties
is presented, results reassert the presence of
Veselago medium.
\end{abstract}

\section{Introduction}

Celestial objects such as Black hole (BH) are abundant in our
universe. In black hole gravitational pull is much stronger that
nothing can escape from its so-called horizon. Relativists suggest
that death of massive stars or collapse of a supergiant star give
birth to black holes. In $1916$, Schwarzschild found the simplest
black hole as an exact solution of field equations \cite{1}.

Plasma is a distinct fourth state of matter having positive and
negatively charged particles that potently interact with
electromagnetic fields \cite{2}. The theory of magnetohydrodynamics
(MHD) is formulated for the study of plasma flow under the influence
of magnetic field. Magnetohydrodynamics in connection with gravity
impressions is recognized as a distinct and authentic theory called
general relativistic magnetohydrodynamics (GRMHD) \cite{3}. Exact
solution of the field equations for a charged object constitute the
Reissner-N{\"o}rdstrom spacetime. The de-Sitter metric is a vacuum
solution of the field equations with a positive cosmological
constant leading to the far future expanding universe
\cite{4}-\cite{6}.

Reissner-N{\"o}rdstrom-de Sitter (RN-dS) metric corresponds to spacetime
solutions characterized by charge, mass and cosmological constant that are assumed
to be most pragmatic for astrophysical applications \cite{7}.
Regge and Wheeler \cite{8}, Zerilli \cite{9} and Gleiser et al. \cite{10} contributed
to establish the stability of non-rotating black holes.
ADM $3+1$ formalism\cite{11} is the most reliable approach for spacetime decomposition in
General relativity, it divides metric into layers of three-dimensional
spacelike hypersurfaces and one-dimensional time. Petterson \cite{12}
concluded that gravitational field is much stronger across the
surface of non-rotating black hole.

Zhang \cite{14} modified the laws of perfect GRMHD in general and
also emphasized on stationary symmetric GRMHD solutions. He
\cite{15} also discuss the rotating black hole dynamics. Buzzi et
al. \cite{16} used ADM $3+1$ split to characterize two fluid plasma
with the help of two local approximations in neighborhood of
Schwarzschild black hole.

Astefanesei et al. \cite{17} worked out Reissner-N{\"o}rdstrom-de Sitter black holes
in context of de-Sitter background to conformal field theory by using static and planar
coordinates. Zhong and Gao \cite{18} investigated particle collisions around cosmological horizon of a Reissner-Nordström
de Sitter solution. Transverse wave propagation
for two fluid plasma has also been acquired for Schwarzschild de-Sitter (SdS) magnetosphere \cite{19}. Sharif and Sheikh \cite{20}-\cite{21}
analyzed plasma modes of cold and isothermal plasmas in the usual medium for
non-rotating and rotating black holes.
Sharif and Rafique \cite{22} presented dispersion modes of
Schwarzschild horizon for the hot plasma in usual medium.

Metamaterials are artificially produced materials with unusual
electromagnetic properties having enormous astrophysical
applications. Russian physicist Victor Veselago introduced such a
metamaterial named Veselago medium that have negative refractive
index, permittivity and permeability simultaneously \cite{23}.
Various scientists \cite{24}-\cite{28} analyzed negative phase
velocity and refractive index of such metamaterial. Ziolkowski and
Heyman \cite{29} developed analytic and numerical solution to
establish wave properties of such medium. Ramakrishna \cite{30}
explained the role of Veselago medium in perfect lensing phenomenon.
After experimental realization of veselago medium, plasma study in Veselago medium
has gained enormous importance in recent years.

Sharif and Mukhtar \cite{31} discussed dispersion relations for
Schwarzschild horizon in unusual medium. Sharif and
Noureen \cite{32} determined dispersion modes of isothermal and hot
plasma across SdS horizon for negative index medium.
Recently, wave behavior around
Reissner-N{\"o}rdstrom spacetime is worked out \cite{34}.

We also use ADM $3+1$ split general line element
\begin{equation}\label{if}
ds^2=-\alpha^2dt^2+\eta_{mn}(dx^m+\beta^mdt)(dx^n+\beta^ndt),
\end{equation}
where lapse function (ratio of fiducial observer (FIDO), i.e, proper
time to universal time, $\frac{d\tau}{dt}$) is expressed by
$\alpha$, $\beta^m$ denotes shift vector in three dimensions and
components of spacelike hypersurfaces are represented by
$\eta_{mn}~(m,n=1,2,3)$. In planar analogue of RN-dS spacetime
$\alpha=\alpha(z),\quad\beta=0,\quad\eta_{mn}=1~(m=n)$ and so metric
turn out to be
\begin{equation}\label{ifr}
ds^2=-\alpha^2(z)dt^2+dx^2+dy^2+dz^2.
\end{equation}
Wave analysis of hot plasma around RN-dS horizon in
Veselago medium is examined in this paper. Since plasma in nature is
highly ionized, so wave analysis of plasma around expanding charged
spacetime is very captivating. The
paper is arranged in the following manner: Fourier analyzed form of
GRMHD equations for rotating black hole are given in section
\textbf{2}. Section \textbf{3} provides numerical
solutions of linearly perturbed Fourier analyzed GRMHD equations for
hot plasma. Comparison and summary is
given in the last section.

\section{Fourier Analyzed GRMHD Equations in a Veselago Medium}

Here, we use basic GRMHD equations (given in appendix) to get
insight of Fourier analyzed GRMHD equations. It is assumed that
plasma flow is two dimensional, i.e., in $xz$-plane. Velocity
$\textbf{V}$ and magnetic field $\textbf{B}$ measured by FIDO are
\begin{eqnarray}\label{9}
\textbf{V}=v(z)\textbf{e}_x+u(z)\textbf{e}_z, \quad
\textbf{B}=B[\xi(z)\textbf{e}_x+\textbf{e}_z].
\end{eqnarray}
Here $B$ denotes constant. The relation between the quantities $\xi,~u$ and $v$ is
\cite{20}
\begin{equation}\label{a}
v=\frac{C}{\alpha}+\xi u,
\end{equation}
where $C$ represents constant of integration. The Lorentz factor,
$\gamma=\frac{1}{\sqrt{1-\textbf{V}^2}}$ turn out to be $\gamma=\frac{1}{\sqrt{1-u^2-v^2}}.$
In order to study the consequences of black hole gravity on plasma, we
implement linear perturbation scheme to the flow variables (pressure $p$, mass density
$\rho$, $\textbf{V}$ and $\textbf{B}$) as follows
\begin{eqnarray}\label{10}
&&\rho=\rho^0+\epsilon\rho=\rho^0+\rho\widetilde{\rho},\quad
p=p^0+\epsilon p=p^0+p\widetilde{p},\nonumber\\
&&\textbf{V}=\textbf{V}^0+\epsilon\textbf{V}=\textbf{V}^0+\textbf{v},~
\textbf{B}=\textbf{B}^0+\epsilon\textbf{B}=\textbf{B}^0+B\textbf{b},
\end{eqnarray}
where $~p, \rho^0, ~\textbf{V}^0$ and $~\textbf{B}^0$ represent the
unperturbed quantities. The linearly perturbed quantities are
denoted by $\epsilon\rho,~\epsilon p,~\epsilon\textbf{V}$ and
$\epsilon\textbf{B}$. Dimensionless
quantities $\widetilde{\rho},~\widetilde{p},~v_x,~v_z,~b_x$ and
$b_z$ are inserted which corresponds to perturbed quantities
\begin{eqnarray}\label{11}
&&\tilde{p}=\tilde{p}(t,z),\quad\tilde{\rho}=\tilde{\rho}(t,z)
,\quad\textbf{v}=\epsilon\textbf{V}=v_x(t,z)\textbf{e}_x
+v_z(t,z)\textbf{e}_z,\nonumber\\
&&\textbf{b}=\frac{\epsilon\textbf{B}}{B}=b_x(t,z)\textbf{e}_x
+b_z(t,z)\textbf{e}_z.
\end{eqnarray}
Application of linear perturbation in perfect GRMHD equations
(\ref{4})-(\ref{8}) and insertion of Eq.(\ref{11}) leads to
component form of linearly perturbed GRMHD equations \cite{32},
provided in the \textbf{Appendix}. Further, to take Fourier analysis
of perturbed GRMHD equations, metric dependence is considered in the
following way
\begin{eqnarray}\label{24}
\tilde{\rho}(t,z)=a_1e^{-\iota\omega t+\iota kz},&\quad&
\tilde{p}(t,z)=a_2e^{-\iota\omega t+\iota kz},\nonumber\\
v_z(t,z)=a_3e^{-\iota\omega t+\iota kz},&\quad&
v_x(t,z)=a_4e^{-\iota\omega t+\iota kz},\nonumber\\
b_z(t,z)=a_5e^{-\iota\omega t+\iota kz},&\quad&
b_x(t,z)=a_6e^{-\iota\omega t+\iota kz},
\end{eqnarray}
where $k$ is $z$-component of the wave vector (wave number), angular
frequency is expressed by $\omega$. To find refractive index and
other plasma properties, we use above wave number. Here dispersion
corresponding to the effects of frequency dependence during wave
propagation. Insertion of Eq.(\ref{24}) and component form of
linearly perturbed GRMHD equations leads to the Fourier analyzed
form,
\begin{eqnarray}\label{25}
&&a_{4}(\alpha'+\iota k\alpha)-a_3 \left\{(\alpha\xi)'+\iota
k\alpha\xi\ \right\}-a_5(\alpha v)'-a_6\{(\alpha
u)'+\iota\omega\nonumber\\
&&+\iota ku\alpha\}=0,\\\label{26}
&&a_5(\frac{-\iota\omega}{\alpha})=0,\\\label{27} &&a_5\iota
k=0,
\\\label{28}&&a_1(\frac{-\iota\omega}{\alpha}\rho)+a_2(\frac{-\iota\omega}{\alpha}p)
+a_3(\rho+p)[\frac{-\iota\omega}{\alpha}\gamma^2u+(1+\gamma^2u^2)\iota k\nonumber\\
&&-(1-2\gamma^2u^2)(1+\gamma^2u^2)\frac{u'}{u}+2\gamma^4u^2vv']
+a_4(\rho+p)\gamma^2[(\frac{-\iota\omega}{\alpha}\nonumber\\
&&+\iota ku)v+u(1+2\gamma^2v^2)v'+2\gamma^2u^2vu']=0,\\
\label{29}&&a_1[\rho\gamma^2u\{(1+\gamma^2v^2)v'+\gamma^2vuu'\}+\gamma^2vu(\rho'+\iota
k\rho)]+a_2[p\gamma^2u\nonumber\\ &&\times\{(1+\gamma^2v^2)v'+\gamma^2vuu'\}+\gamma^2vu(p'+\iota kp)]+a_3[(\rho+p)\gamma^2
\{(1\nonumber\\ &&+2\gamma^2u^2)(1+2\gamma^2v^2)v'+(\frac{-\iota\omega}{\alpha}+\iota
ku)\gamma^2vu-\gamma^2v^2v'+2\gamma^2(1+\nonumber\\ &&2\gamma^2u^2)uvu'\}
+\gamma^2v(1+2\gamma^2u^2)(\rho'+p')-\frac{B^2u}{4\pi\alpha}(\xi\alpha)'+\frac{\xi
B^2}{4\pi}(\frac{\iota\omega}{\alpha}\nonumber\\ &&-\iota
ku)]+a_4[(\rho+p)\gamma^4u\{(1+4\gamma^2v^2)
uu'+4vv'(1+\gamma^2v^2)\}+(\rho+\nonumber\\ &&
p)\gamma^2(1+\gamma^2v^2)(\frac{-\iota\omega}{\alpha}+\iota
ku)+\gamma^2u(1+2\gamma^2v^2)(\rho'+p')+\frac{B^2}{4\pi}(\frac{u\alpha'}{\alpha}-\nonumber\\&&
\frac{\iota\omega}{\alpha}+\iota
ku)]-a_6\frac{B^2}{4\pi\alpha}[\alpha
uu'+\alpha'(1+u^2)+(1+u^2)\iota k\alpha]=0,
\\\label{30}
&&a_1[\rho\gamma^2\{a_z+(1+\gamma^2u^2)uu'+\gamma^2u^2vv'\}+\gamma^2u^2(\rho'+\iota
k\rho)]+a_2[p\gamma^2\{a_z\nonumber\\
&&+(1+\gamma^2u^2)uu'+\gamma^2u^2vv'\}+(1+\gamma^2u^2)(p'+\iota kp)]+a_3[(\rho+p)\gamma^2
\{(1\nonumber\\
&&+\gamma^2u^2)(\frac{-\iota\omega}{\alpha}+\iota
ku)+u'(1+\gamma^2u^2)(1+4\gamma^2u^2)+2u\gamma^2(a_z
+(1+2\gamma^2u^2)\nonumber\\ &&\times
vv')\}+2\gamma^2u(1+\gamma^2u^2)(\rho'+p')+\frac{\xi
B^2u}{4\pi\alpha}(\xi\alpha)'-\frac{\xi^2
B^2}{4\pi}(\frac{\iota\omega}{\alpha}-\iota ku)]\nonumber\\
&&+a_4[(\rho+p)\gamma^4\{(\frac{-\iota\omega}{\alpha}+\iota
ku)uv+u^2v'(1+4\gamma^2v^2)+2v(a_z+  \nonumber
\end{eqnarray}
\begin{eqnarray}
&&(1+2\gamma^2u^2)uu')\}+2\gamma^4u^2v(\rho'+p')+\frac{\xi
B^2}{4\pi}(\frac{\iota\omega}{\alpha}-\iota ku)-\frac{\xi B^2u\alpha'}{4\pi\alpha}]+\nonumber\\
&& a_6[\frac{B^2}{4\pi\alpha}\{-(\xi\alpha)'
+\alpha'\xi-u\xi(u\alpha'+u'\alpha)\}+\frac{\xi
B^2}{4\pi}(1+u^2)\iota k]=0,
\\\label{31}&&a_1\{(\frac{-\iota\omega}{\alpha}\gamma^2+\iota ku
\gamma^2+2u\gamma^2a_z+\gamma^2u')\rho+u\rho'\gamma^2\}
+a_2\{(\frac{\iota\omega}{\alpha}(1-\gamma^2)\nonumber\\
&&+\iota ku\gamma^2+2\gamma^2ua_z+\gamma^2u')p+u\gamma^2p'\}+a_3\gamma^2\{(\rho'+p')+2(2\gamma^4uu'+\nonumber\\
&&a_z+2\gamma^2u^2a_z)(\rho+p)+(1+2\gamma^2u^2)(\rho+p)\iota
k+\frac{\xi B^2}{4\pi\alpha}(\xi
u-v)\iota\omega+\alpha\xi'\}\nonumber\\
&&+a_4[2(\rho+p)\gamma^2\{(u
v'+2uva_z+u'v)+uv\iota
k\}+\frac{B^2}{4\pi\alpha}(v-u\xi)\iota\omega-\alpha\xi']+\nonumber\\
&&
a_6[\frac{-B^2}{4\pi\alpha}\{(v^2+u^2)\xi+\xi
v(\xi v+u)\iota\omega\}-\alpha\xi'u+\iota k\alpha(v-u\xi)]=0.
\end{eqnarray}

\section{Hot plasma dispersion modes}

The equation of state is used to obtain the numerical solution of
Fourier analyzed form for rotating environment. For hot
plasma, it is given by \cite{14}
\begin{eqnarray}\label{3}
\mu=\frac{\rho+p}{\rho_0}.
\end{eqnarray}
The specific enthalpy $\mu$ in Eq.(\ref{3}) is
non-constant, so affects the perturbed GRMHD equations
(\ref{25})-(\ref{31}) for hot plasma around the RN-dS black horizon.
In the following subsections, we discuss the hot plasma case
in the magnetized and non-magnetized backgrounds.

\subsection{Non-magnetized Background}

In non-magnetized environment, $B=0=\xi$ leads to $a_5=0=a_6$, in
the Fourier analyzed GRMHD equations (\ref{28})-(\ref{31})
\cite{32}. In order to solve Fourier analyzed equations numerically,
we assume $\mu=\sqrt{\frac{1+\alpha^{2}}{2}}$ and
$\rho=p=\frac{\mu}{2}$, $\alpha=\frac{z}{2r}$,  $r=r_+ + r_h$, with
$r_+=\frac{z}{2(M+\sqrt{M^2+Q^2})}$, $\frac{Q^2}{M^2}=0.7$,
$u=v=-\frac{1}{\sqrt{z^{2}+2}}$, and $
\gamma=\frac{1}{\sqrt{1-u^2-v^2}}= \frac{\sqrt{z^{2}+2}}{z}$, where
$r_h\thickapprox
2M\left(1+\frac{4M^{2}}{l^{2}}+...\right)\backsimeq\zeta 2.948km$,
$1\leqslant\zeta\leqslant1.5$ and $M\thicksim1M_{\bigodot}$
\cite{19}.

The region under consideration is $4\leq z\leq10$ with event horizon
at $z=0$. The region $0\leq z\leq4$ is ignored because wave propagation can not
be viewed graphically and result are not interesting
in this region.  A dispersion relation is formed from determinant of
coefficients of the corresponding Fourier analyzed equations for
non-magnetized plasma. Real part of dispersion relation evolve
quartic equation in $k$, given as
\begin{equation}\label{36}
H_1(z)k^4+H_2(z,\omega)k^3+H_3(z,\omega)k^2+H_4(z,\omega)k+H_5(z,\omega)=0.
\end{equation}
Above equation has two real and two imaginary roots. Imaginary part
of dispersion relation is cubic in $k$ as follows
\begin{equation}\label{37}
D_1(z)k^3+D_2(z,\omega)k^2+D_3(z,\omega)k+D_4(z,\omega)=0.
\end{equation}
One real and two complex roots are originated from imaginary part.

Wave number $k$ can be used to obtain refractive index whose change
$\frac{dn}{d\omega}$ decides whether the dispersion is anomalous or
normal, equivalently the region of greater phase velocity than the
group velocity corresponds to normal dispersion, otherwise anomalous
\cite{33}. Wave phenomenon around event horizon can be beneficial
only for those values of $k$ for which waves propagate in opposite
direction of the horizon and dispersion is normal. However, no
information can be extracted about magnetosphere for waves that
disperse anomalously.

It is to be mentioned here that dispersion relations are solved by
using software \textit{Mathematica}. There are few roots for which
change in refractive index does not show three dimensional plot but
its values are shown on axis. Some complicated expressions that
mathematica could not plot in the form of graph causes this.

Figures \textbf{1}-\textbf{3} provide three
dimensional view of wave properties of non-magnetized hot plasma around
RN-dS horizon, that are further explained in table I.
\begin{center}
Table I. Dispersion and direction of waves
\begin{tabular}{|c|c|}
 \hline
 In direction of horizon  &  Fig. $1, 3$   \\ \hline
   Moving away from horizon  &  Fig. $2$  \\\hline
 Normal dispersion  &  Fig. $3$ \\  \hline
  Anomalous dispersion  &  Fig. $1, 2$ \\  \hline
     & Fig. $1$, $4.3\leq z\leq7.9,0\leq\omega\leq 9$ \\
     Increasing $n$ with  & Fig. $2$, $4\leq z\leq 9.2, 0\leq\omega\leq10$ \\
     increasing $z$
      & Fig. $3$, $7.5\leq z\leq9, 1\leq\omega\leq 8$\\\hline
\end{tabular}
\end{center}
\begin{figure}
\begin{tabular}{cc}
\epsfig{file=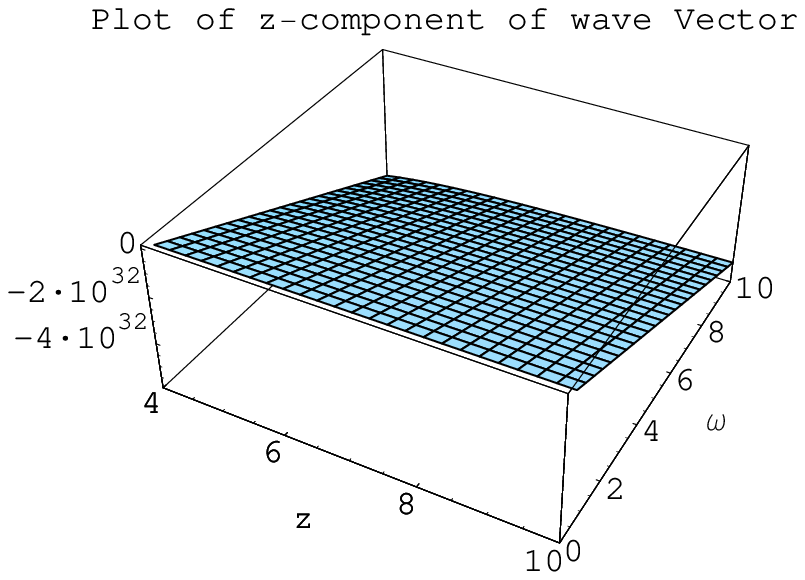,width=0.34\linewidth}
\epsfig{file=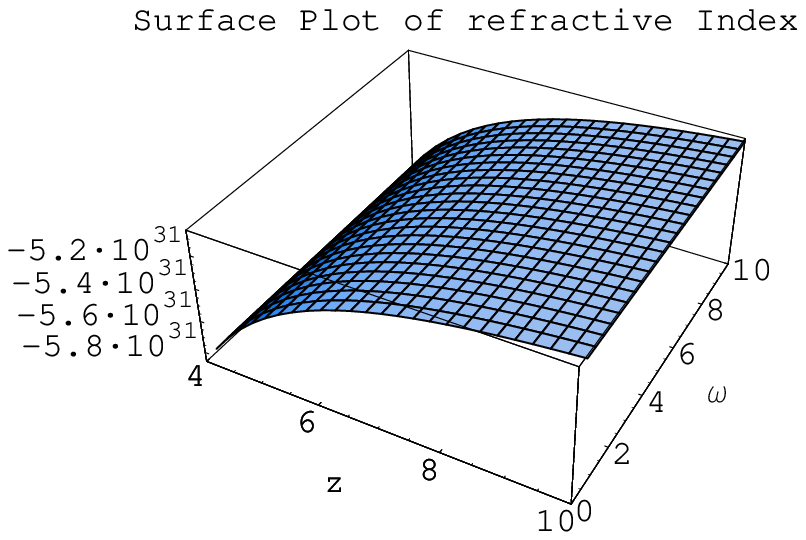,width=0.34\linewidth}\\
\epsfig{file=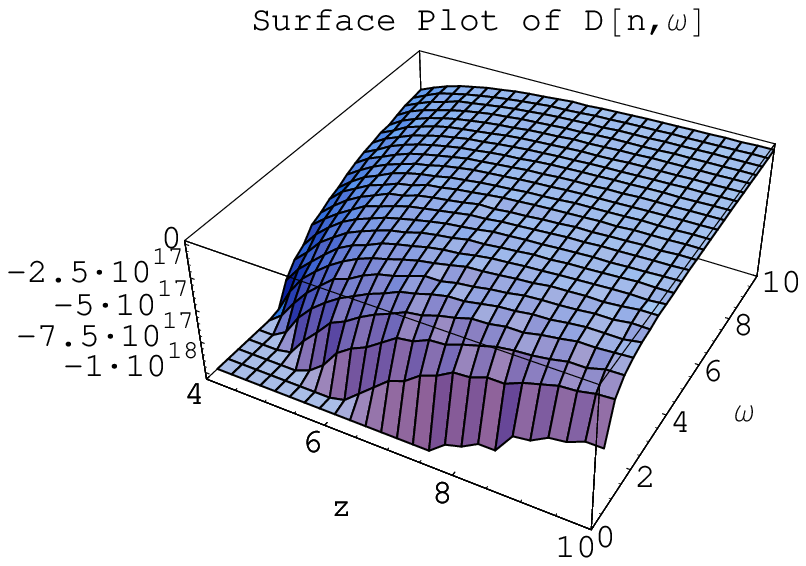,width=0.34\linewidth}
\epsfig{file=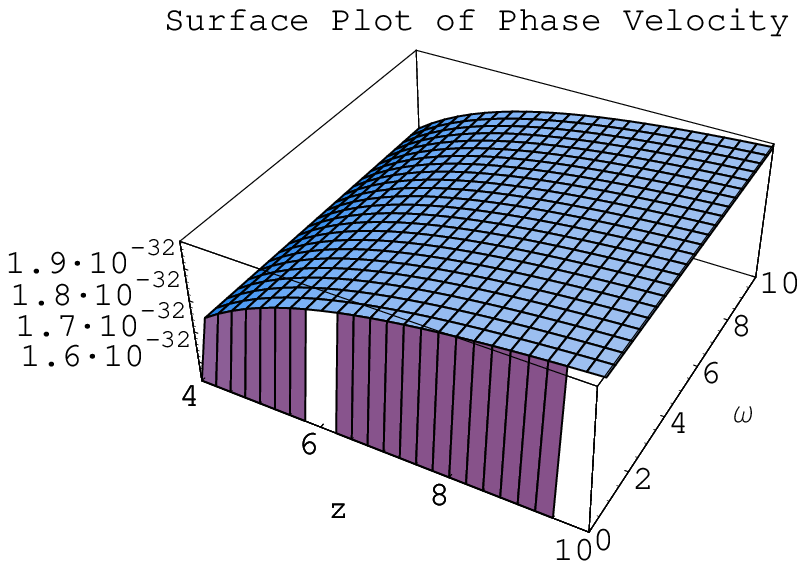,width=0.34\linewidth}
\epsfig{file=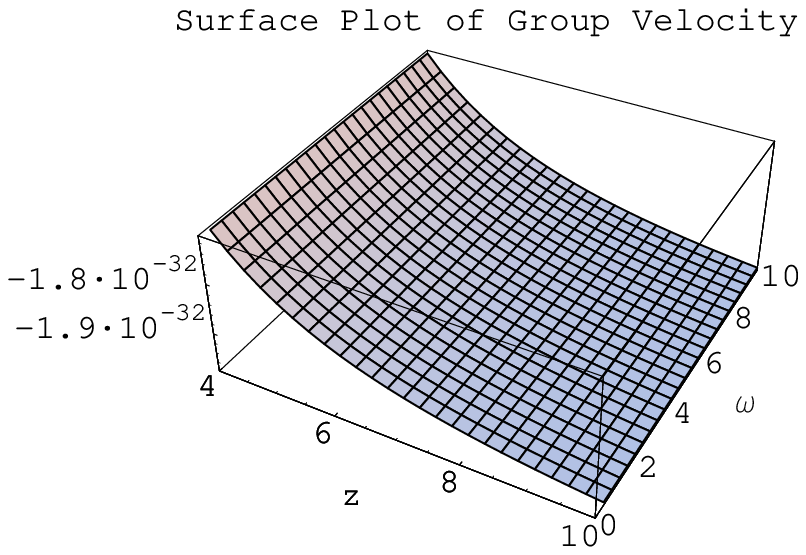,width=0.34\linewidth}\\
\end{tabular}
\caption{Dispersion is normal and anomalous in the region}
\begin{tabular}{cc}\\
\epsfig{file=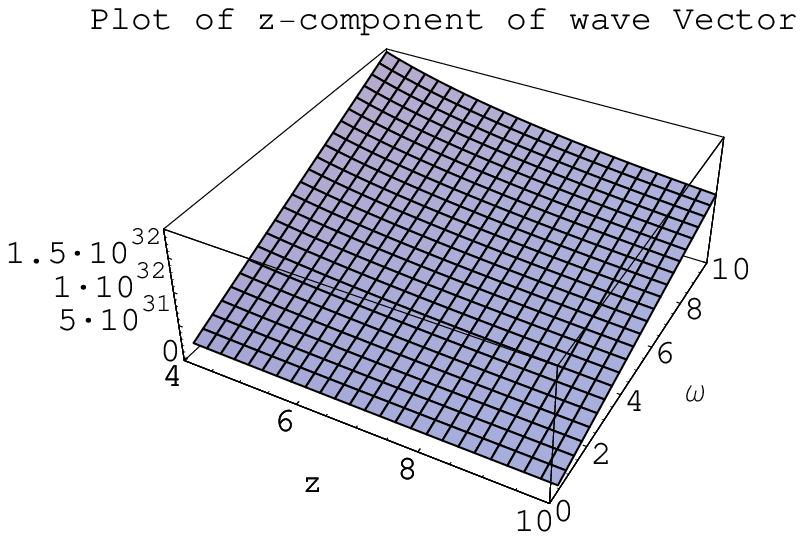,width=0.34\linewidth}
\epsfig{file=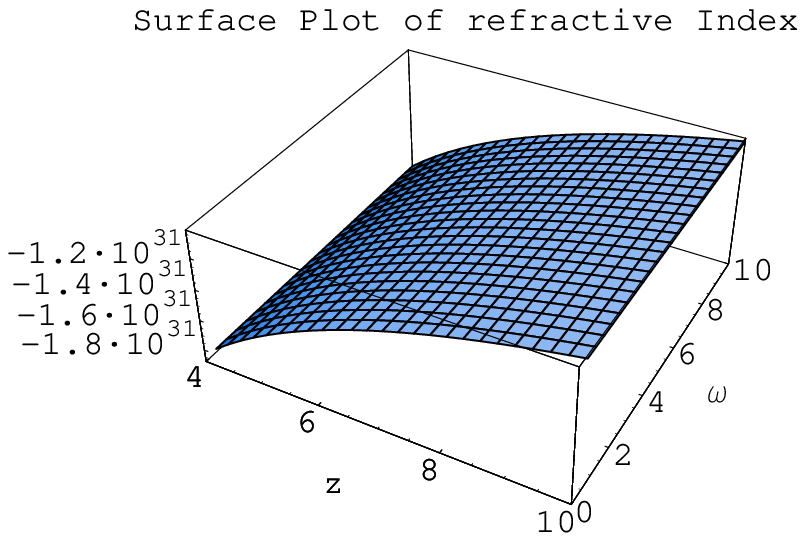,width=0.34\linewidth}\\
\epsfig{file=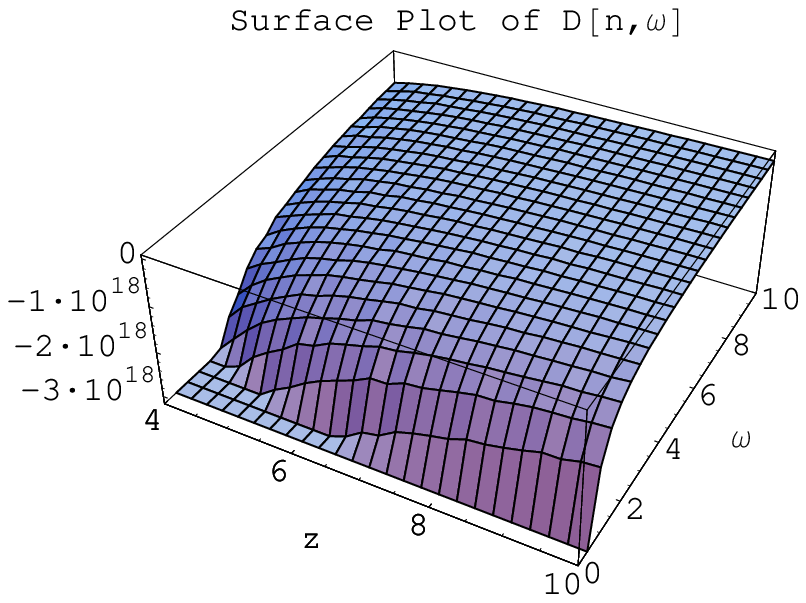,width=0.34\linewidth}
\epsfig{file=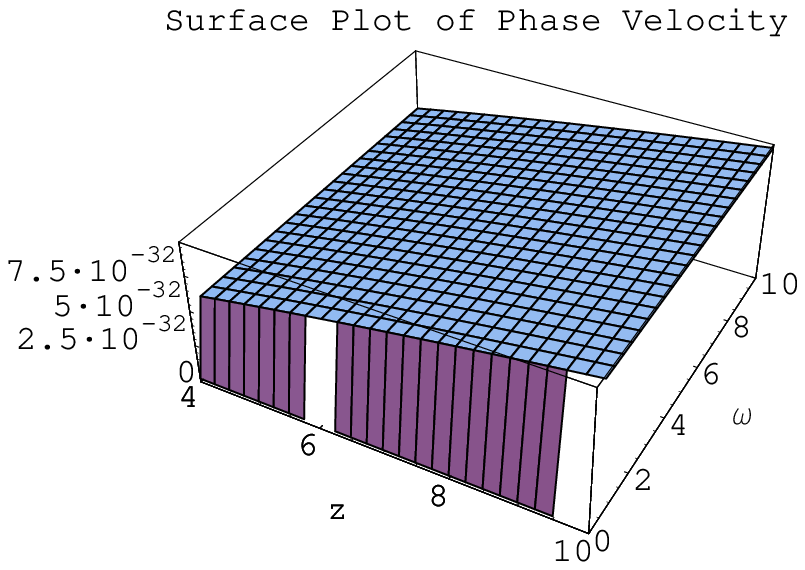,width=0.34\linewidth}
\epsfig{file=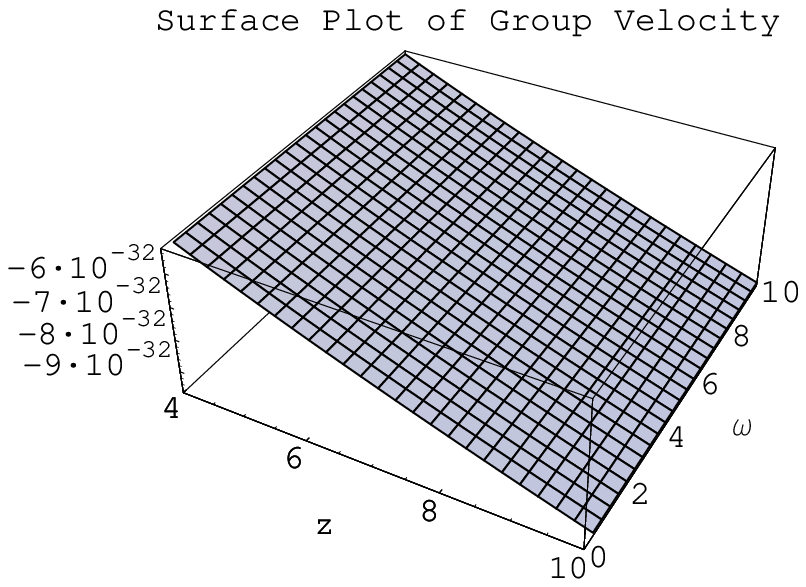,width=0.34\linewidth}\\
\end{tabular}
\caption{Normal as wells as anomalous dispersion occur at random
points in the region.}
\end{figure}
\begin{figure}
\begin{tabular}{cc}
\epsfig{file=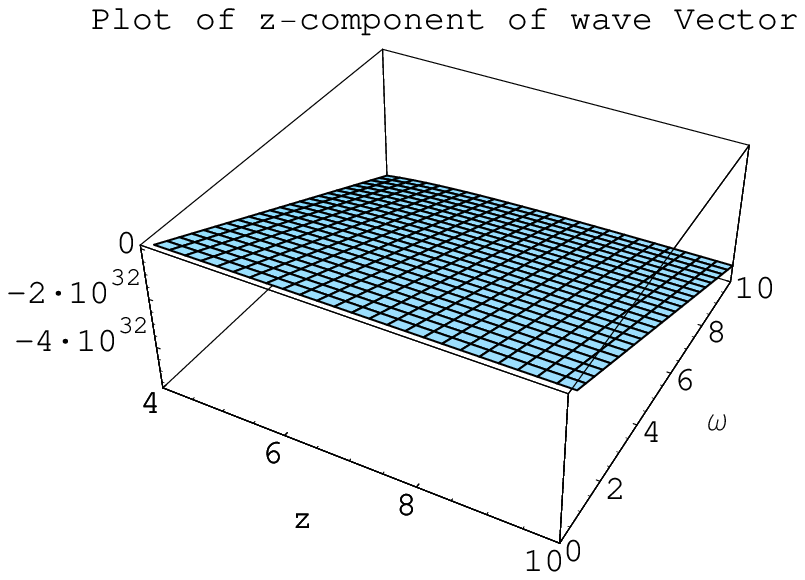,width=0.34\linewidth}
\epsfig{file=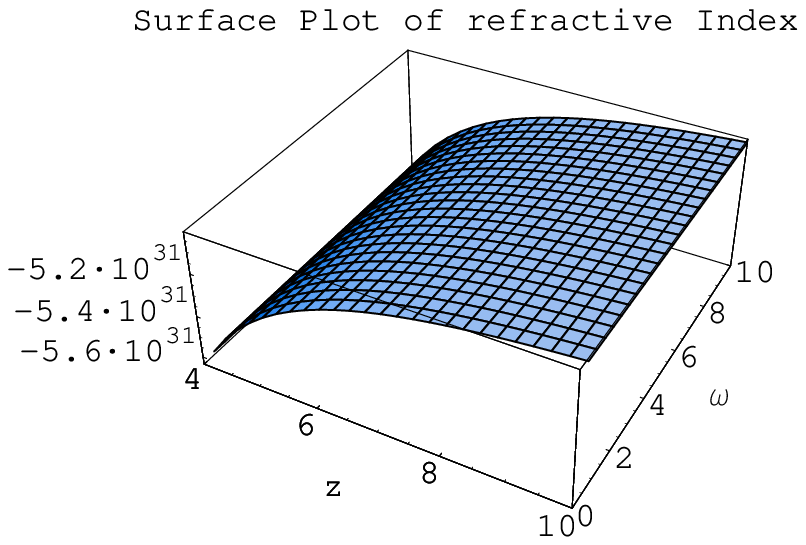,width=0.34\linewidth}\\
\epsfig{file=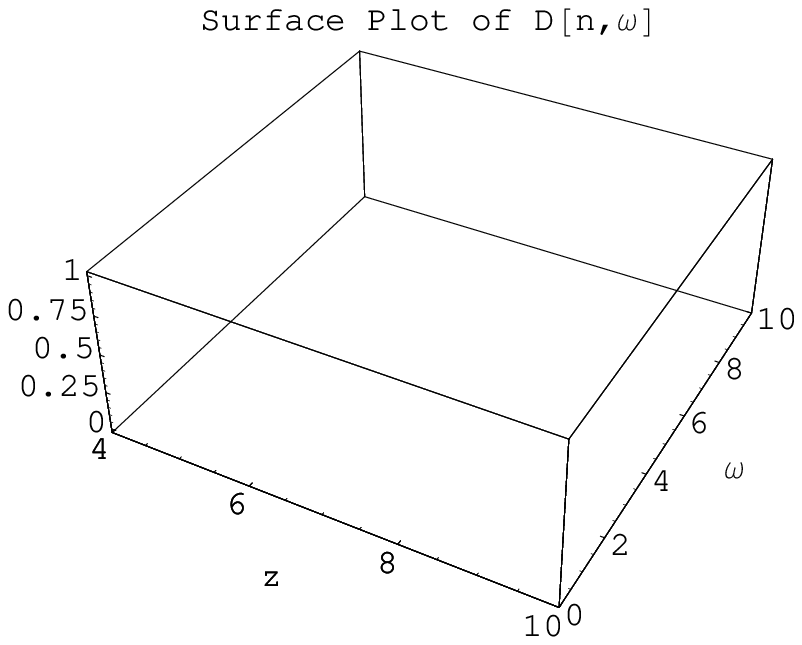,width=0.34\linewidth}
\epsfig{file=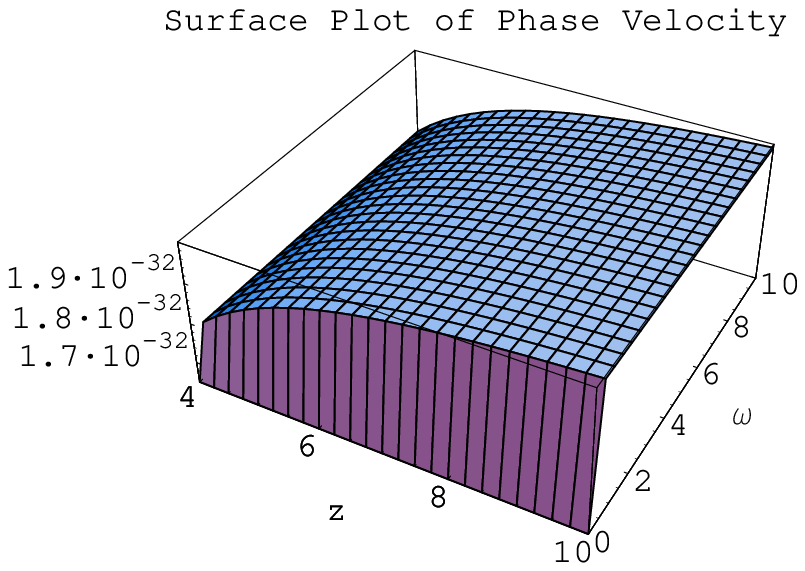,width=0.34\linewidth}
\epsfig{file=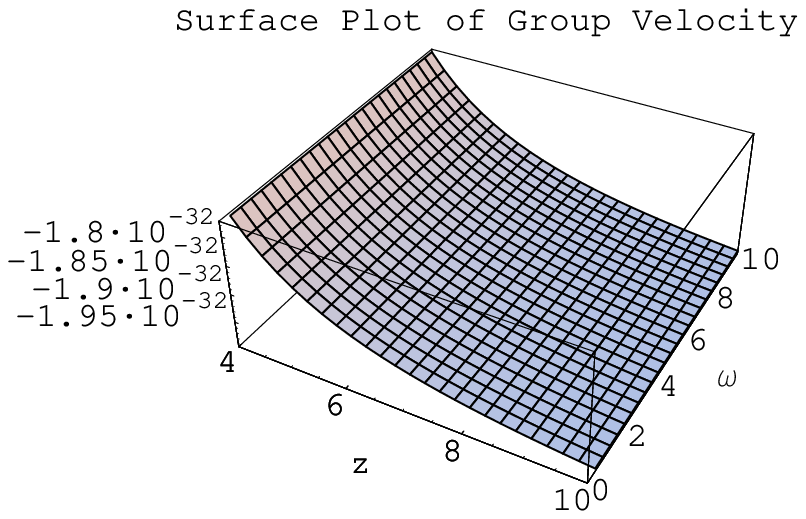,width=0.34\linewidth}\\
\end{tabular}
\caption{Normal and anomalous dispersion of waves is observed.}
\end{figure}

\subsection{Plasma flow in Magnetized Background}

For rotating magnetized plasma flow, the perturbed Fourier analyzed
GRMHD equations are given by Eqs.(\ref{25})-(\ref{31}). In this
case, flow is two dimensional hence \textbf{V} and \textbf{B} lie in
$xz$-plane. For numerical solutions velocity, lapse function and
specific enthalpy are the same as for non-magnetized background.
Here, magnetic field is assumed  to be $\frac{B^{2}}{4\pi}=2$ with
$u=v$. By putting $C=1$ in Eq.(\ref{a}), we have
$\xi=1+\frac{\sqrt{2+z^{2}}}{z}$. Region for wave analysis is
considered to be $4\leq z\leq10,~ 0\leq\omega\leq 10$ and in this case
Eqs.(\ref{26})-(\ref{27}) yields $a_{5}=0$. The determinant of
Fourier analyzed form generates dispersion relation whose real part
is as follows
\begin{equation}\label{38}
D_1(z)k^4+D_2(z,\omega)k^3+D_3(z,\omega)k^2+D_4(z,\omega)k+D_5(z,\omega)=0.
\end{equation}
Four real roots originates from the real part of elaborated in Figures \textbf{4}-\textbf{7}. Imaginary part is given by
\begin{eqnarray}\label{39}
&&E_1(z)k^5+E_2(z,\omega)k^4+E_3(z,\omega)k^3+E_4(z,\omega)k^2+E_5(z,\omega)k\nonumber\\
&&+E_6(z,\omega)=0.
\end{eqnarray}
yield five roots,
three real and two complex described by Figures
\textbf{8}-\textbf{10}. Plasma modes for magnetized background are tabulated
in Table II and III.
\begin{center}
Table II. Refractive index and direction of waves
\begin{tabular}{|c|c|}
 \hline
  In direction of horizon  &  Fig. $4, 5, 6, 8, 9$   \\ \hline
   Moving away from horizon  &  Fig. $7, 10$  \\\hline
 Normal dispersion  &  Fig. $4$ \\  \hline
  Anomalous dispersion  &  Fig. $5$ \\  \hline
     & Fig. $5$, $6.1\leq z\leq9.9,1\leq\omega\leq 8$ \\
     &  Fig. $6$, $4.6\leq z\leq8.6,1\leq\omega\leq 9.6$ \\
     Increasing $n$ with  & Fig. $7$, $4\leq z\leq 8.2, 0\leq\omega\leq9$ \\ increasing $z$
      & Fig. $8$, $4.3\leq z\leq9.4, 1\leq\omega\leq 9$\\
      & Fig. $10$, $4\leq z\leq6.8, 1\leq\omega\leq 9$\\\hline
    Decreasing $n$ with  & Fig. $4$, $5.7\leq z\leq
10,0\leq\omega\leq 4$ \\ increasing $z$ &  Fig. $9$, $4.3\leq z\leq4.9,1.3\leq\omega\leq 9.1$
\\\hline
\end{tabular}
\end{center}
\begin{figure}
\begin{tabular}{cc}\\
\epsfig{file=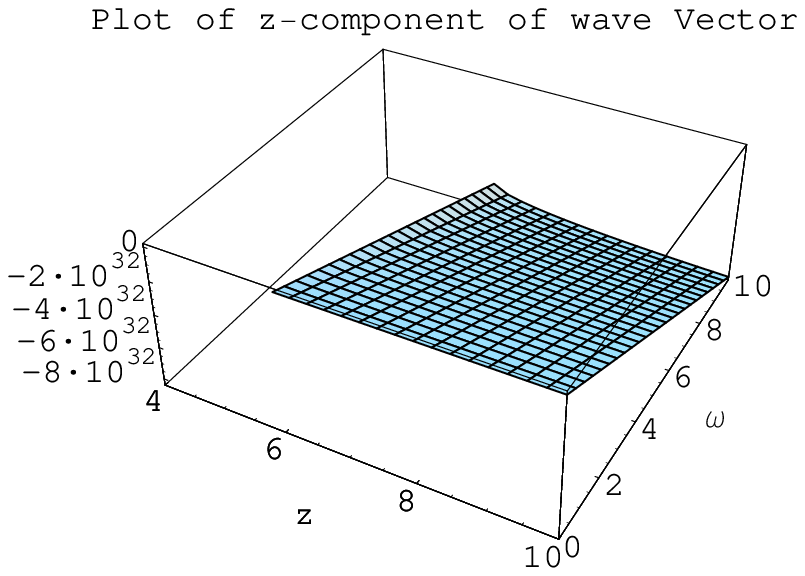,width=0.34\linewidth}
\epsfig{file=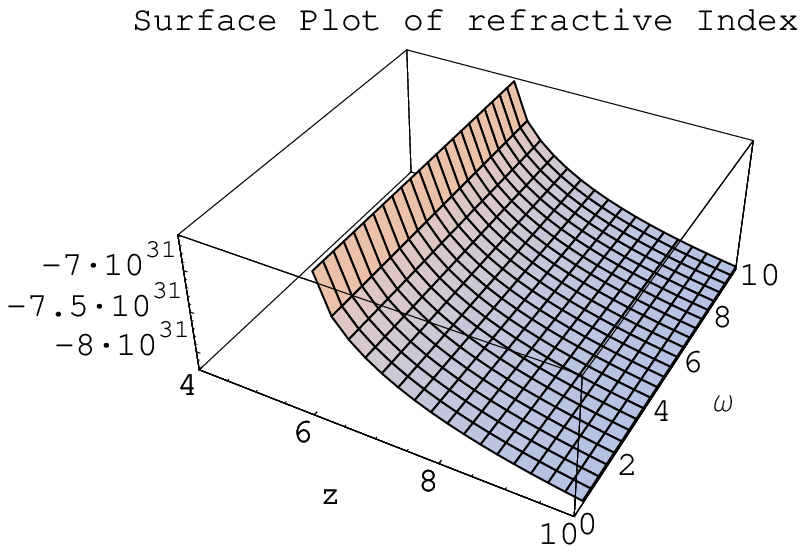,width=0.34\linewidth}\\
\epsfig{file=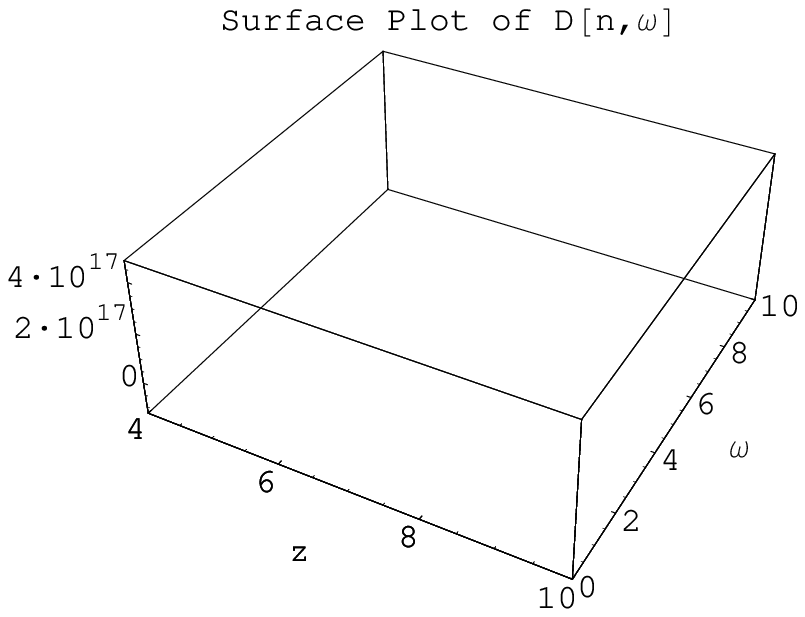,width=0.34\linewidth}
\epsfig{file=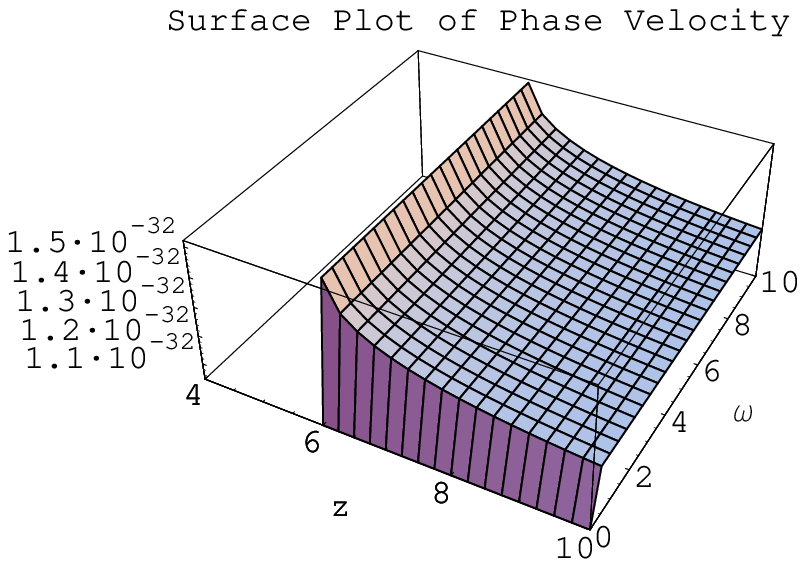,width=0.34\linewidth}
\epsfig{file=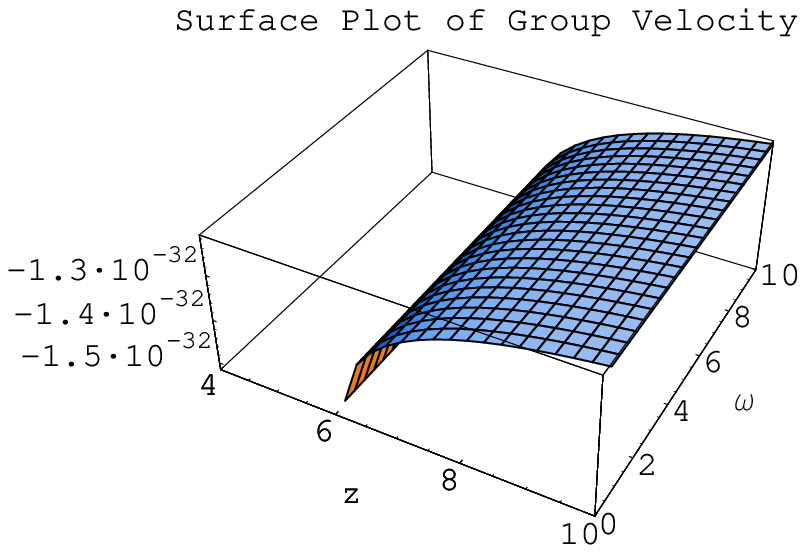,width=0.34\linewidth}\\
\end{tabular}
\caption{Random points of normal and anomalous dispersion are found
in the region.}
\end{figure}
\begin{figure}
\begin{tabular}{cc}
\epsfig{file=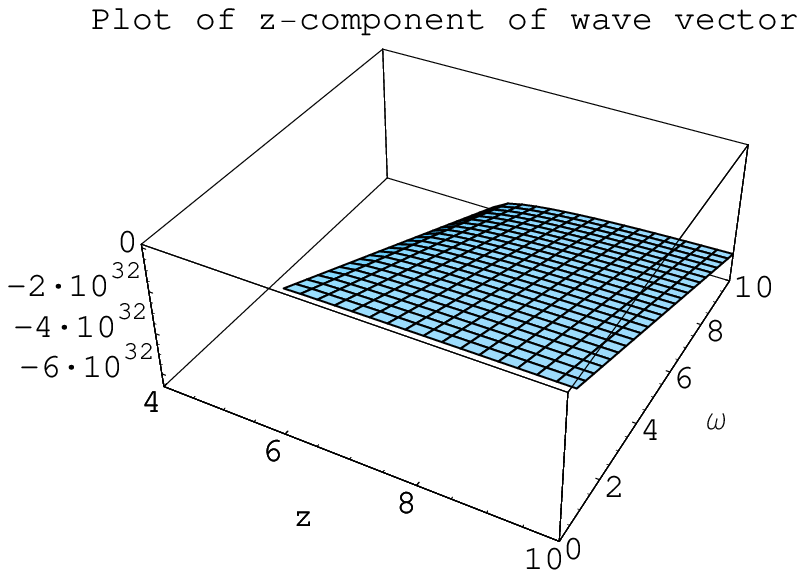,width=0.34\linewidth}
\epsfig{file=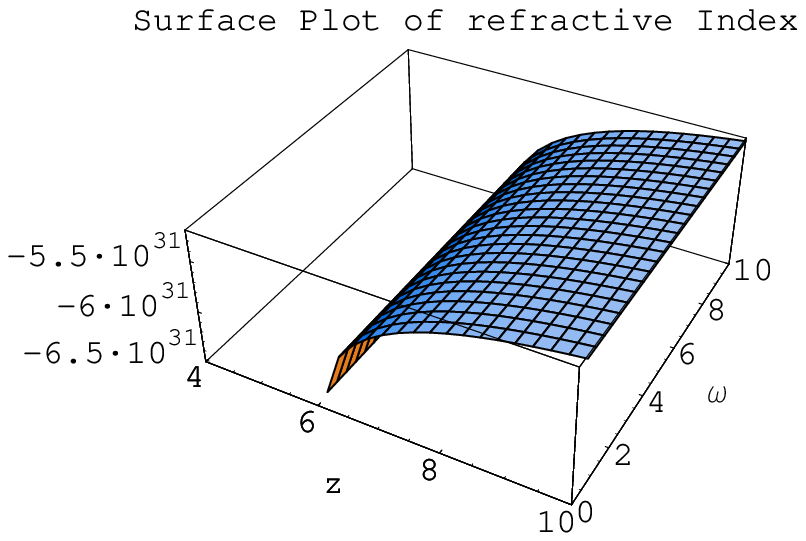,width=0.34\linewidth}\\
\epsfig{file=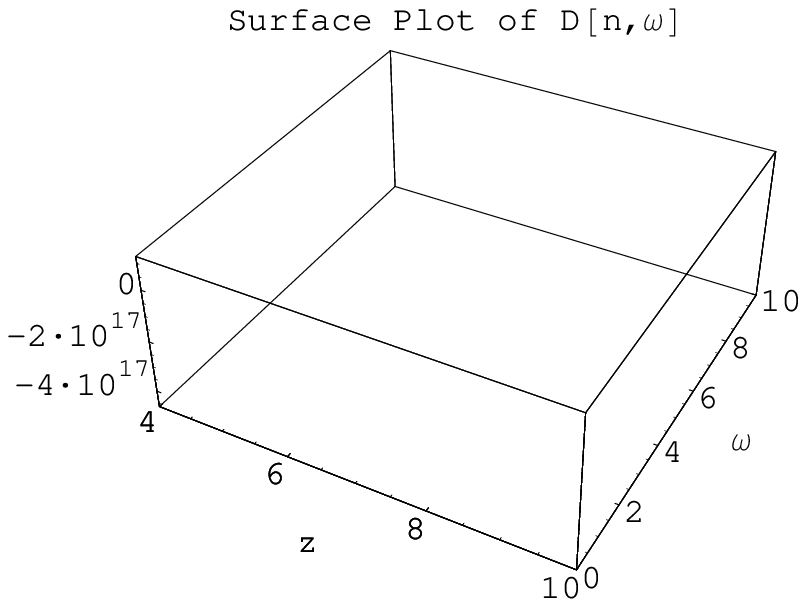,width=0.34\linewidth}
\epsfig{file=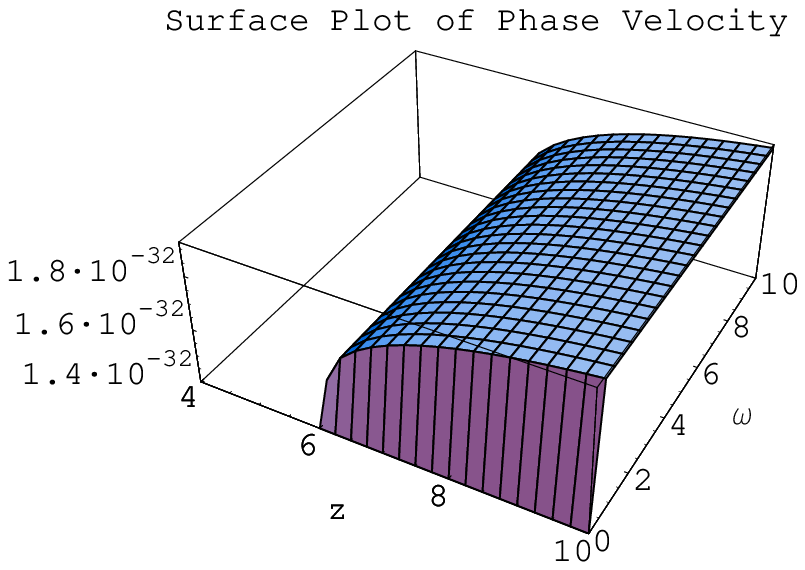,width=0.34\linewidth}
\epsfig{file=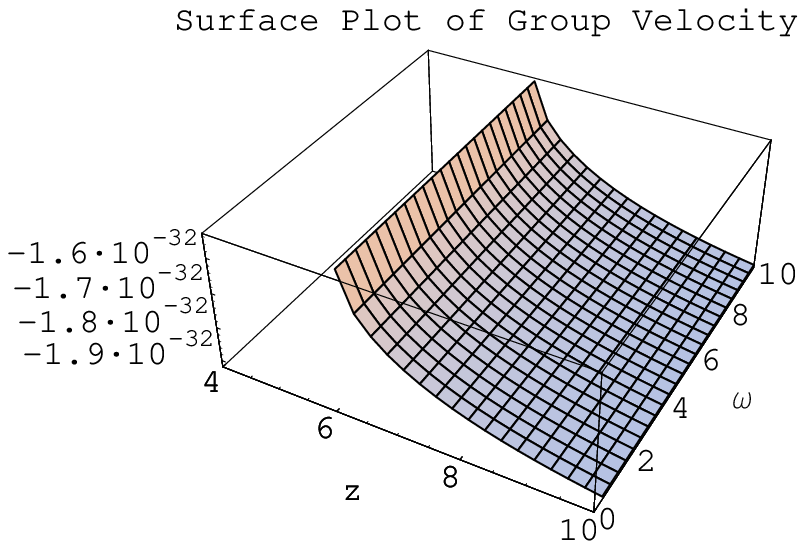,width=0.34\linewidth}\\
\end{tabular}
\caption{Whole region admits normal dispersion.}
\end{figure}
\begin{figure}
\begin{tabular}{cc}\\
\epsfig{file=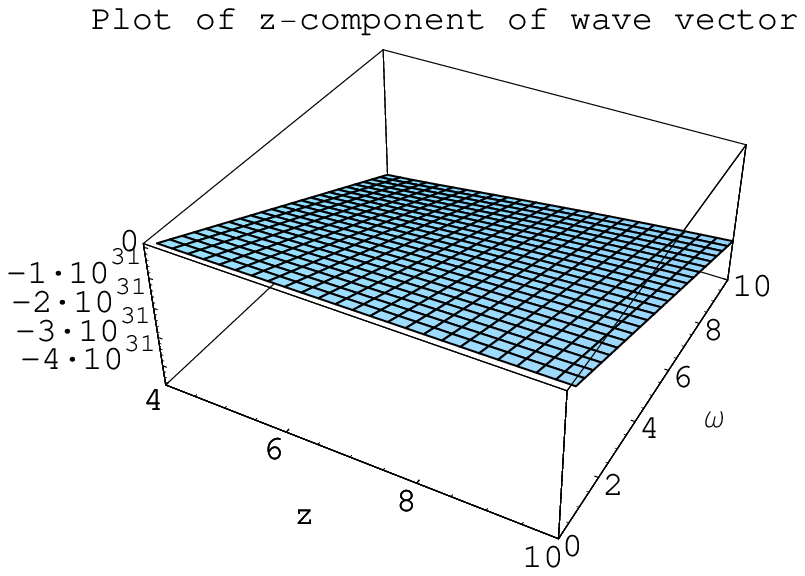,width=0.34\linewidth}
\epsfig{file=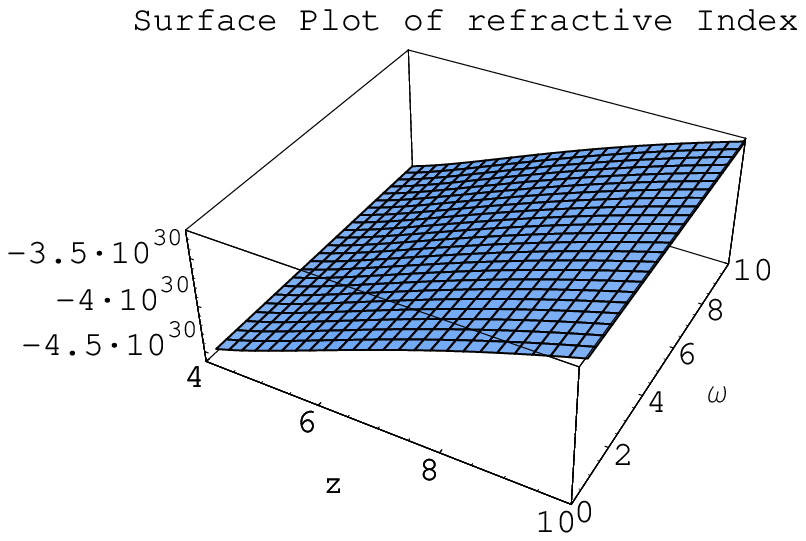,width=0.34\linewidth}\\
\epsfig{file=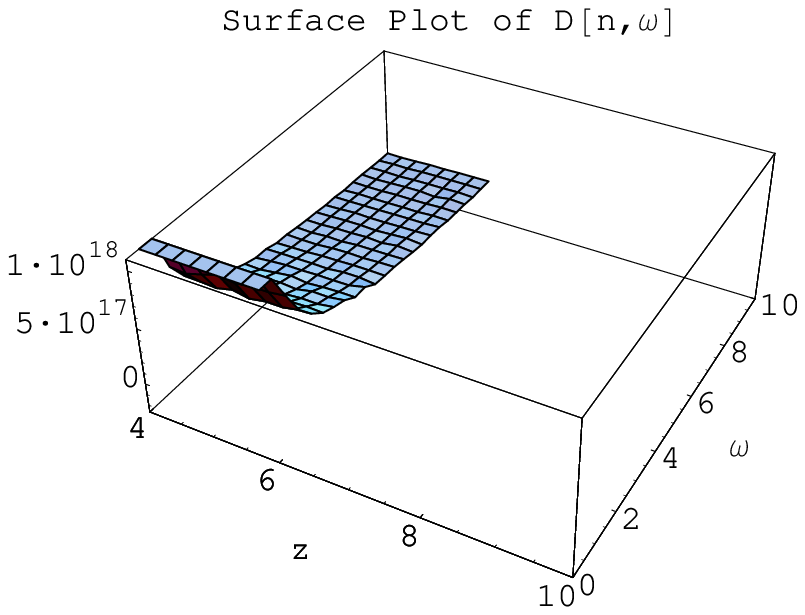,width=0.34\linewidth}
\epsfig{file=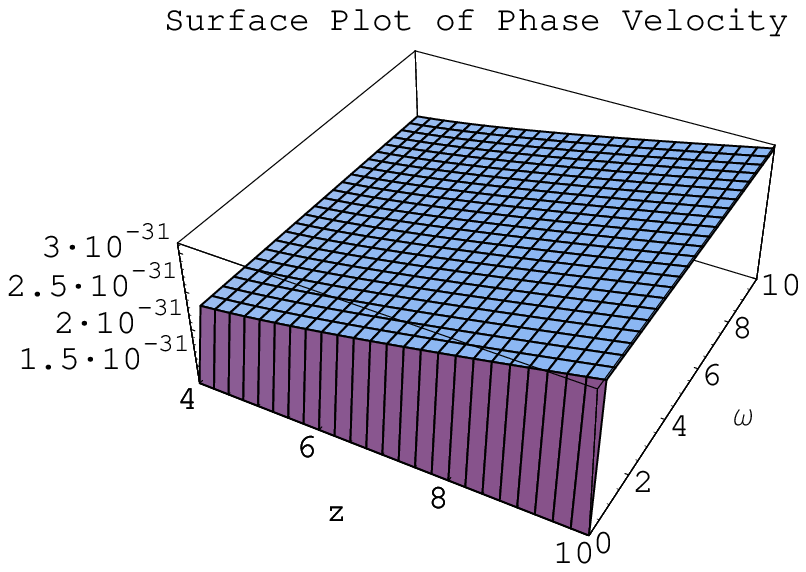,width=0.34\linewidth}
\epsfig{file=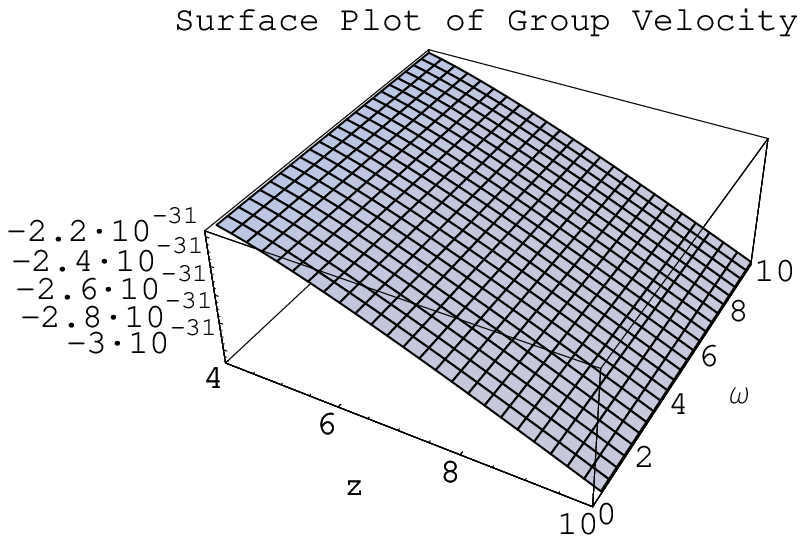,width=0.34\linewidth}\\
\end{tabular}
\caption{Normal and anomalous dispersion of waves is observed.}
\begin{tabular}{cc}
\epsfig{file=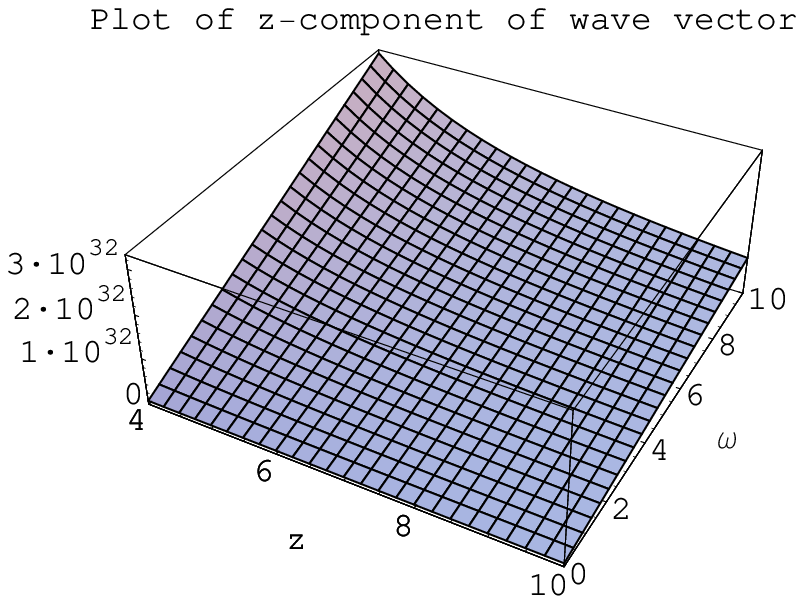,width=0.34\linewidth}
\epsfig{file=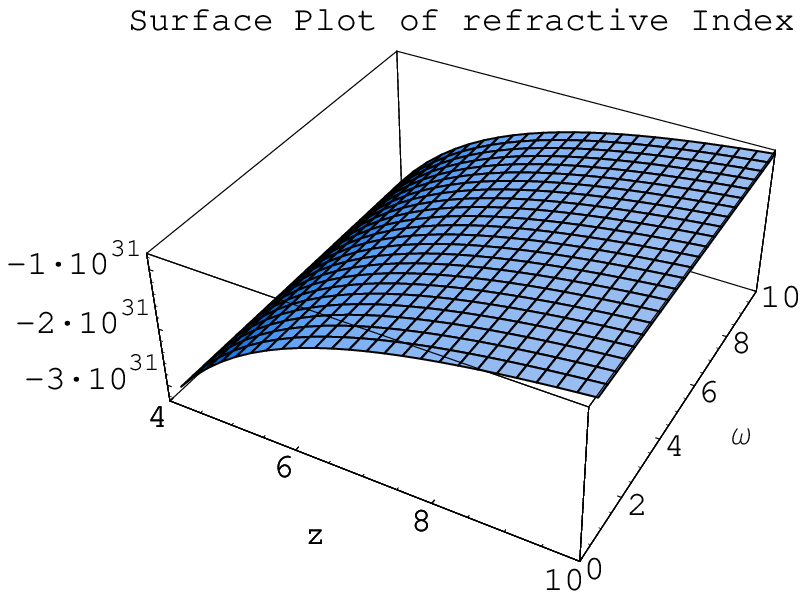,width=0.34\linewidth}\\
\epsfig{file=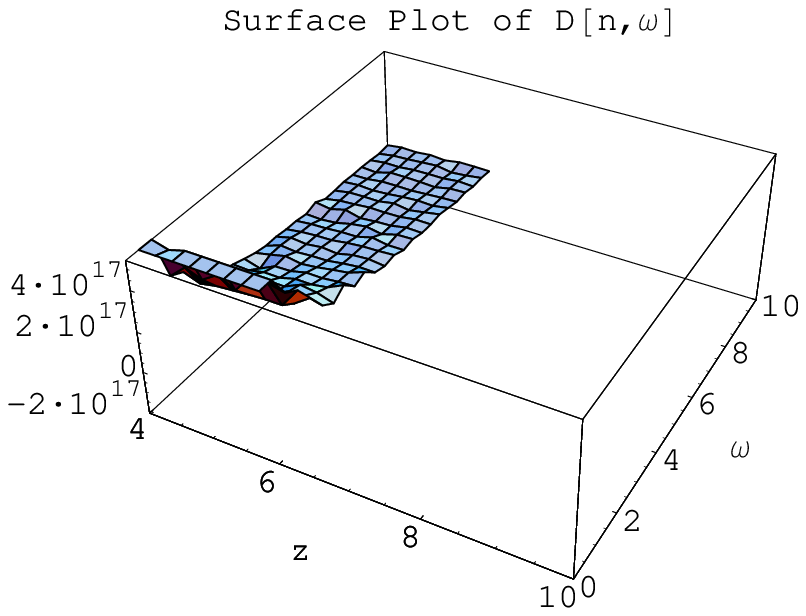,width=0.34\linewidth}
\epsfig{file=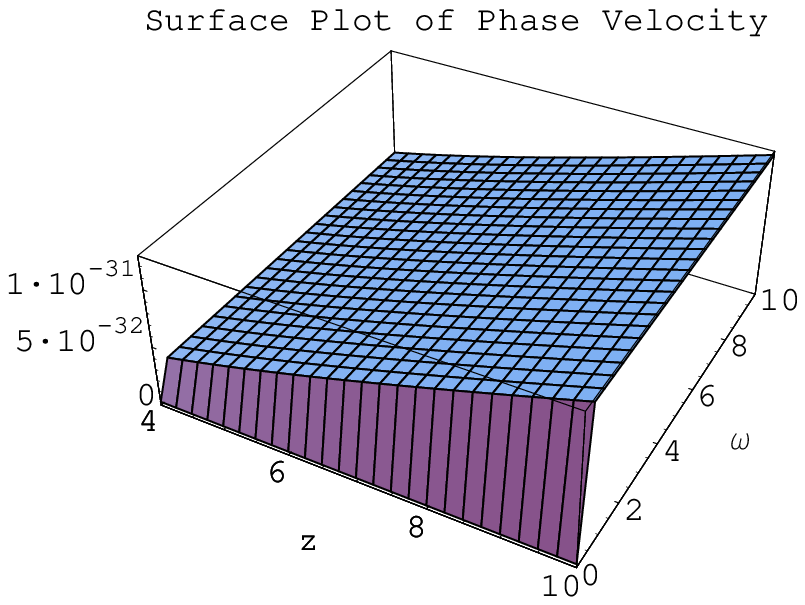,width=0.34\linewidth}
\epsfig{file=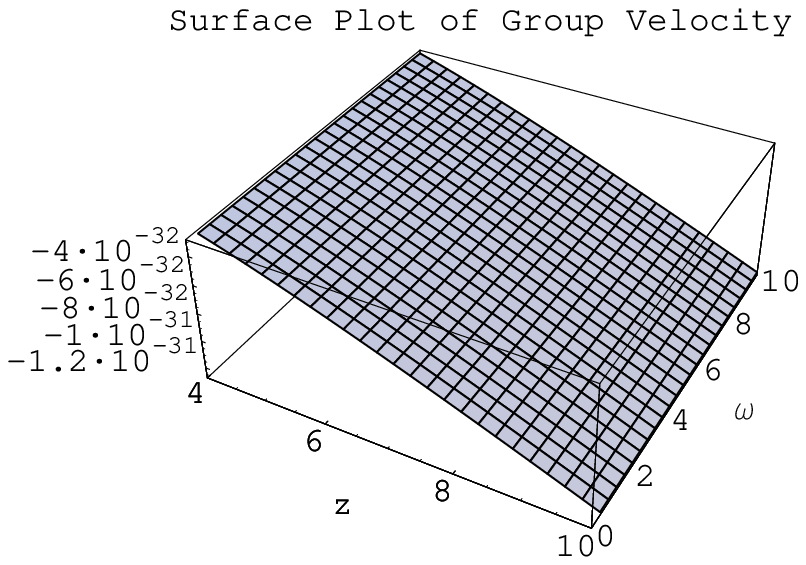,width=0.34\linewidth}\\
\end{tabular}
\caption{Waves disperse normally in the whole region.}
\end{figure}
\begin{figure}
\begin{tabular}{cc}\\
\epsfig{file=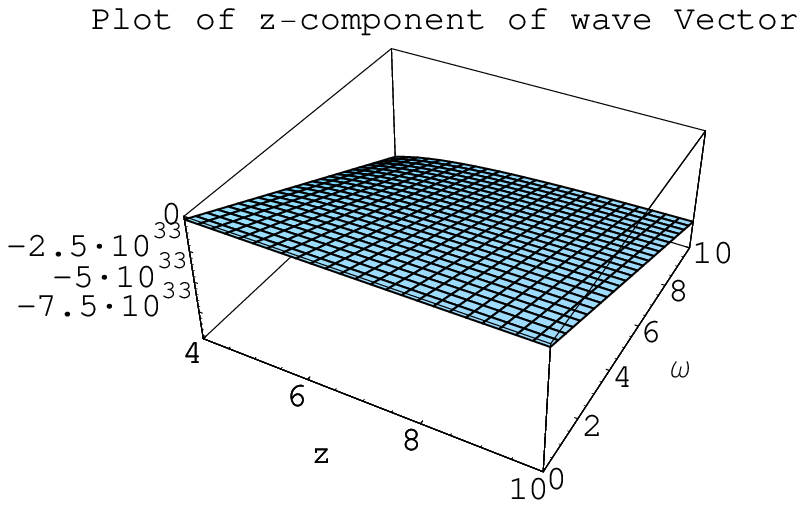,width=0.34\linewidth}
\epsfig{file=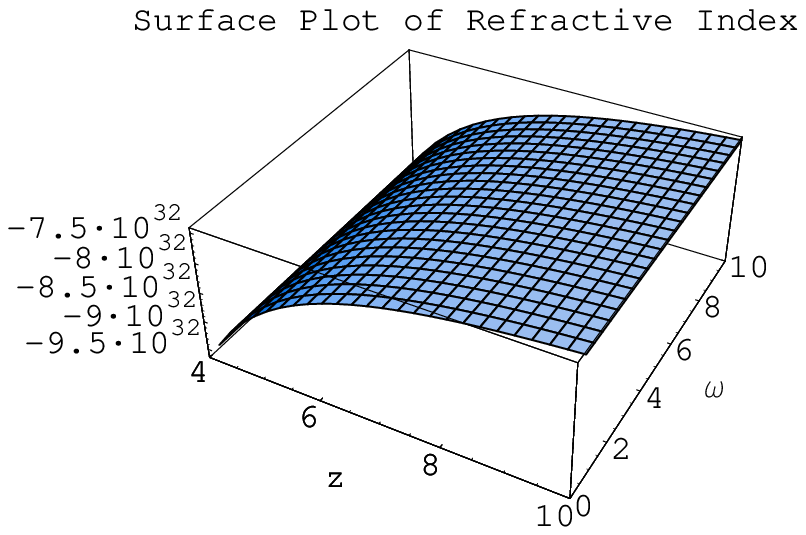,width=0.34\linewidth}\\
\epsfig{file=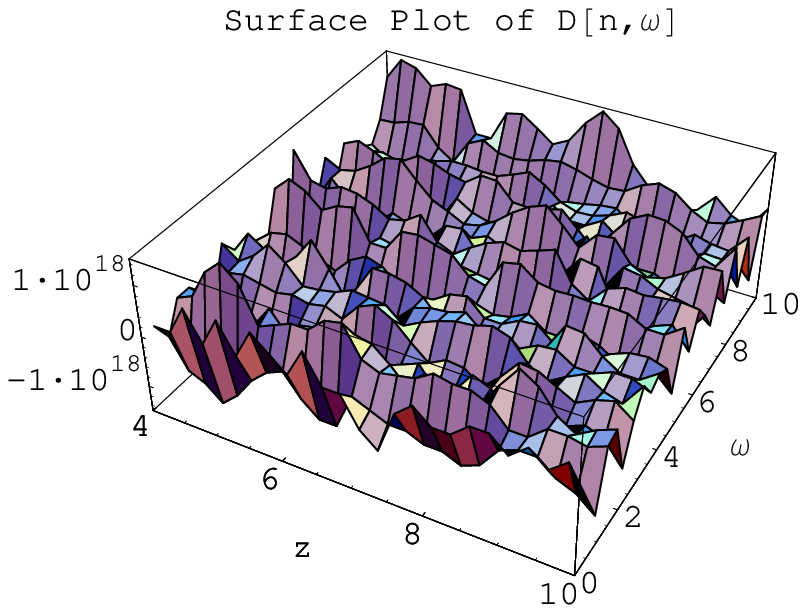,width=0.34\linewidth}
\epsfig{file=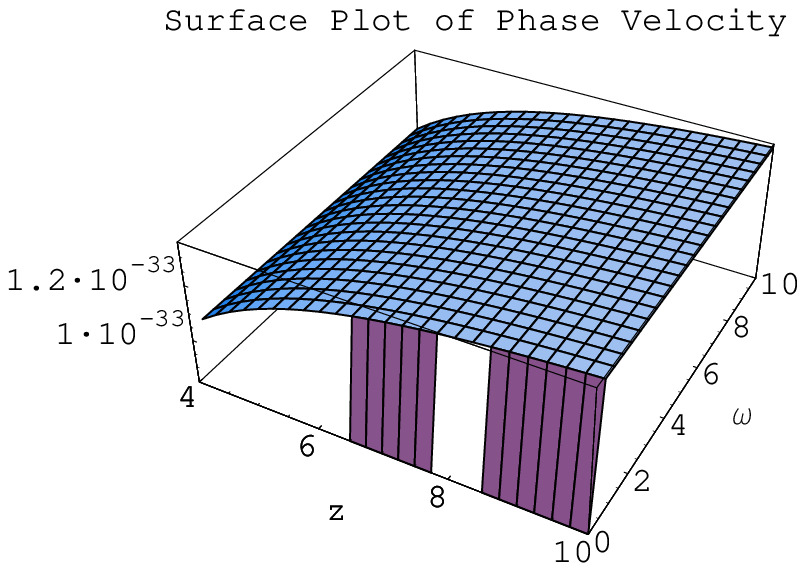,width=0.34\linewidth}
\epsfig{file=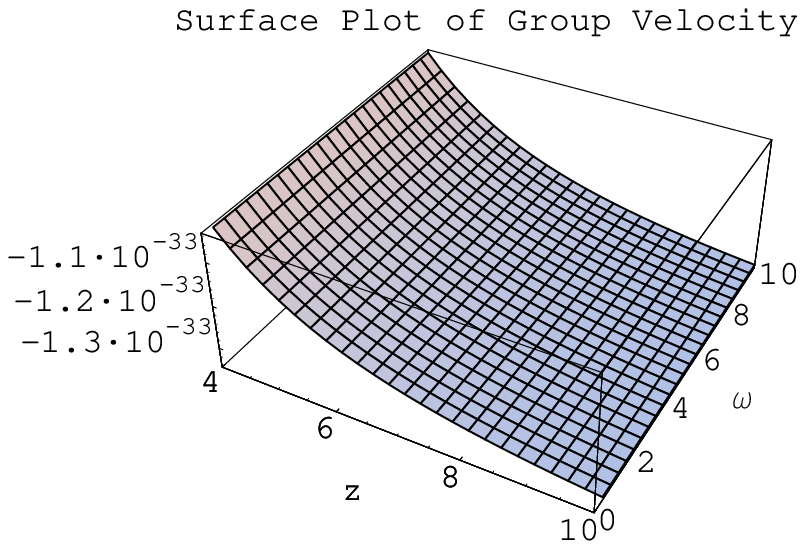,width=0.34\linewidth}\\
\end{tabular}
\caption{Waves exhibit normal as well as normal dispersion at random
points.}
\end{figure}
\begin{figure}
\begin{tabular}{cc}
\epsfig{file=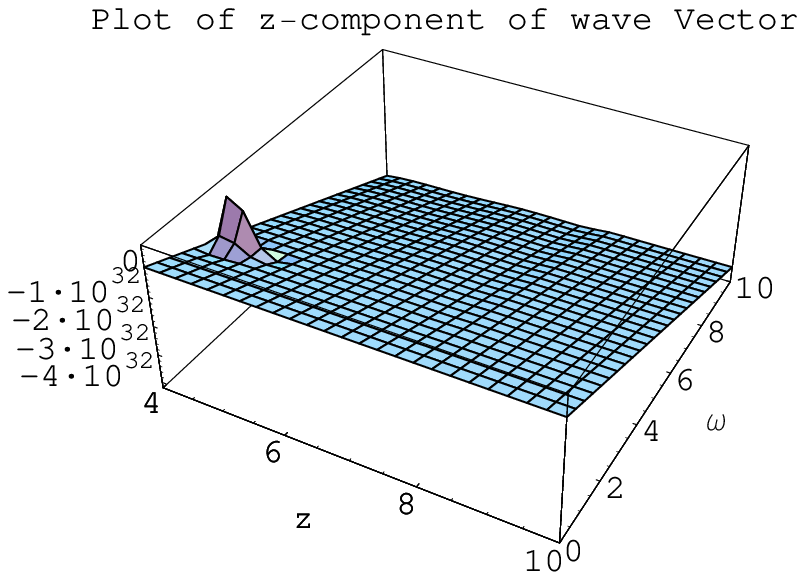,width=0.34\linewidth}
\epsfig{file=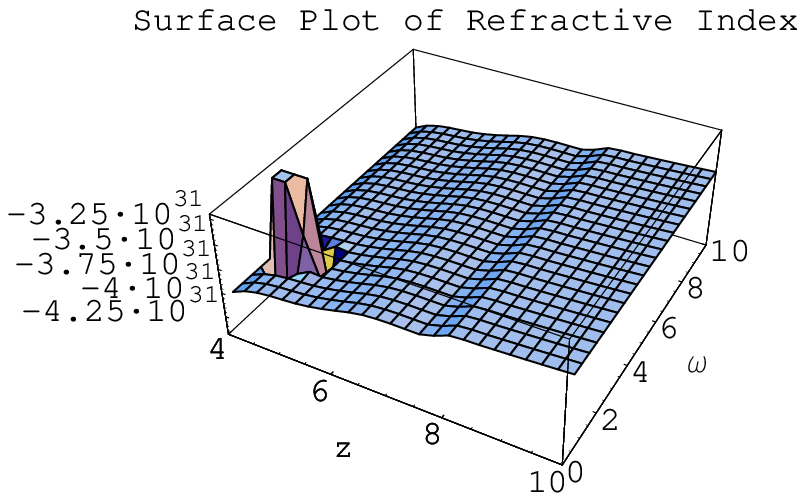,width=0.34\linewidth}\\
\epsfig{file=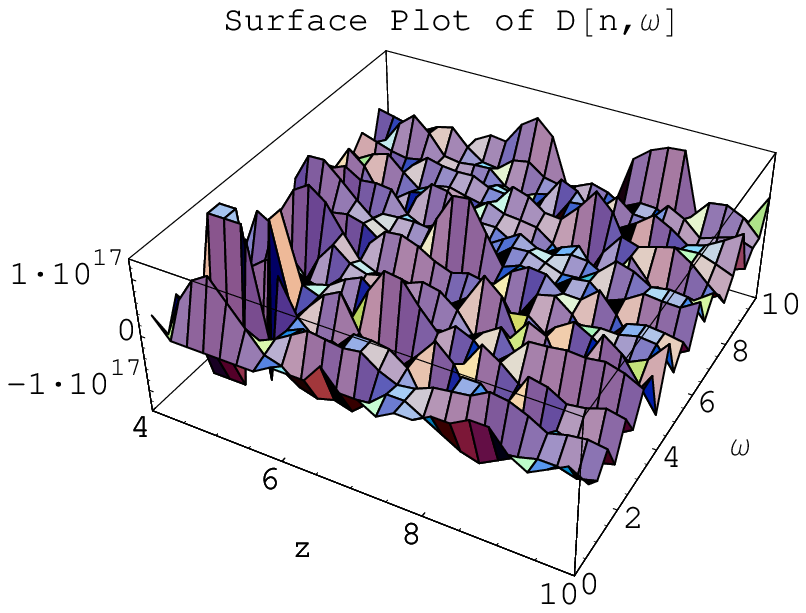,width=0.34\linewidth}
\epsfig{file=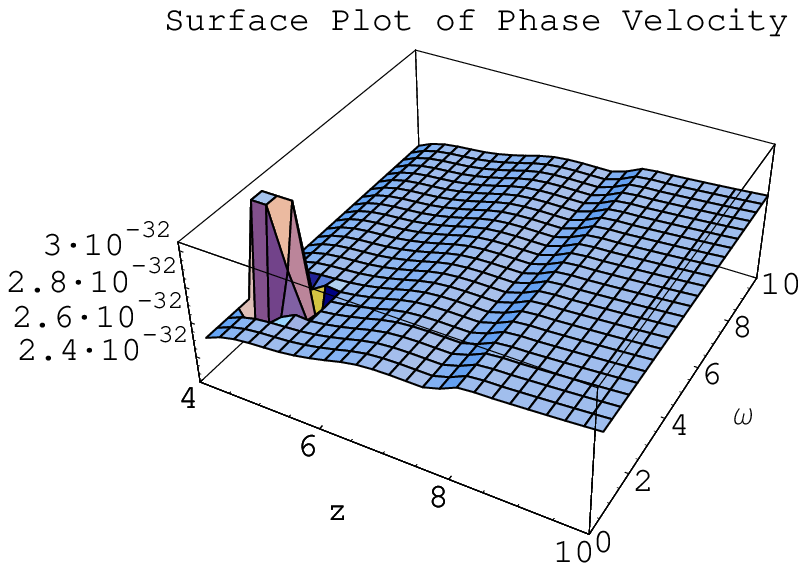,width=0.34\linewidth}
\epsfig{file=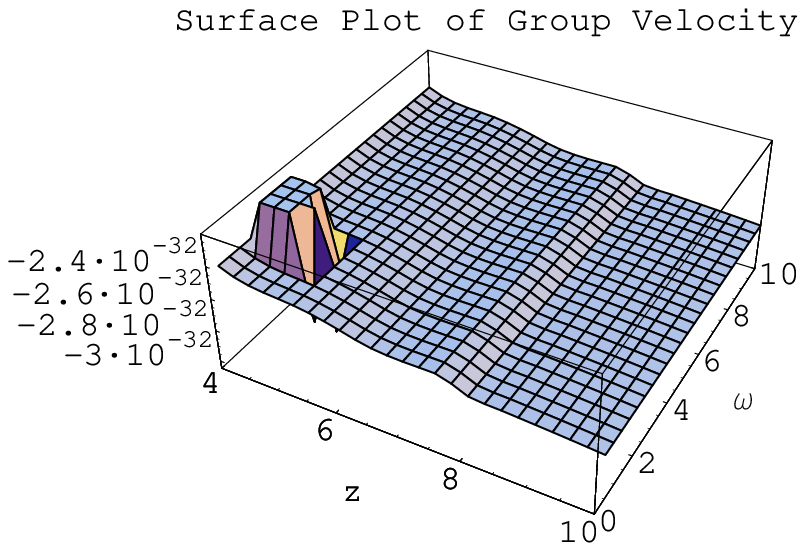,width=0.34\linewidth}\\
\end{tabular}
\caption{Waves disperse normally in the whole region.}
\end{figure}
\begin{figure}
\begin{tabular}{cc}\\
\epsfig{file=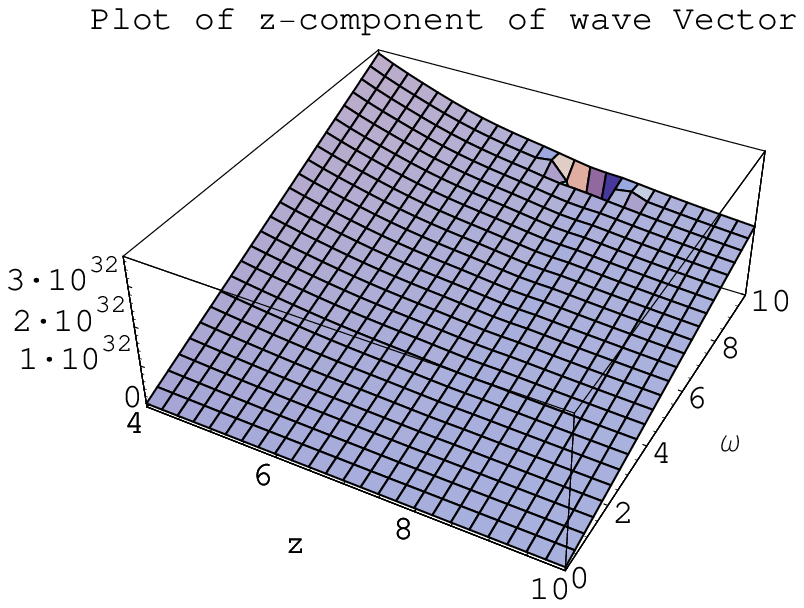,width=0.34\linewidth}
\epsfig{file=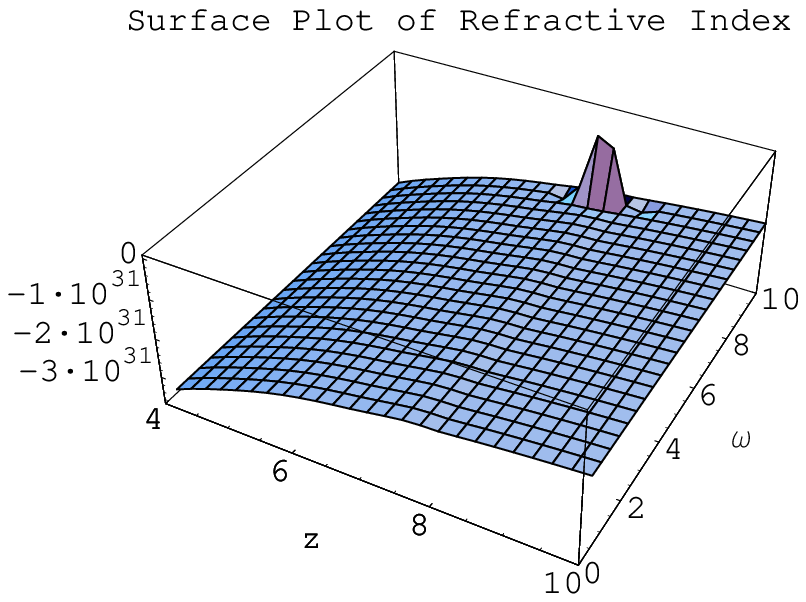,width=0.34\linewidth}\\
\epsfig{file=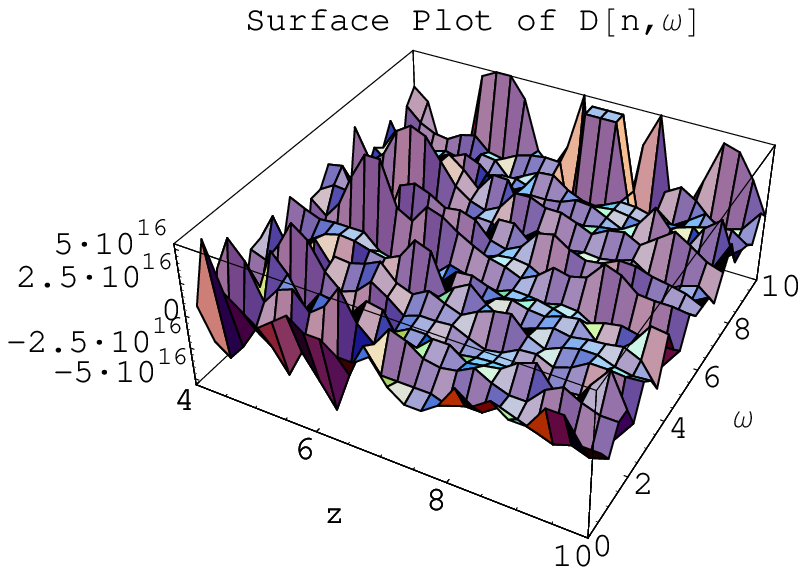,width=0.34\linewidth}
\epsfig{file=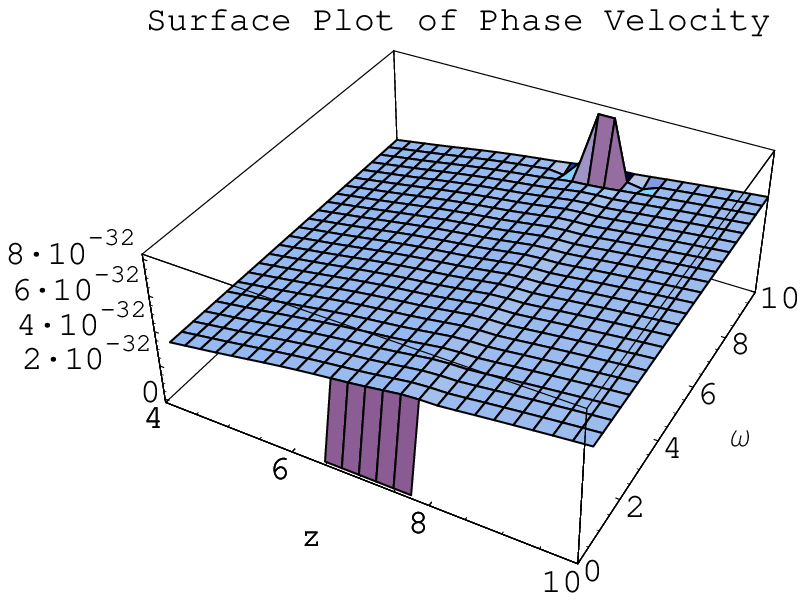,width=0.34\linewidth}
\epsfig{file=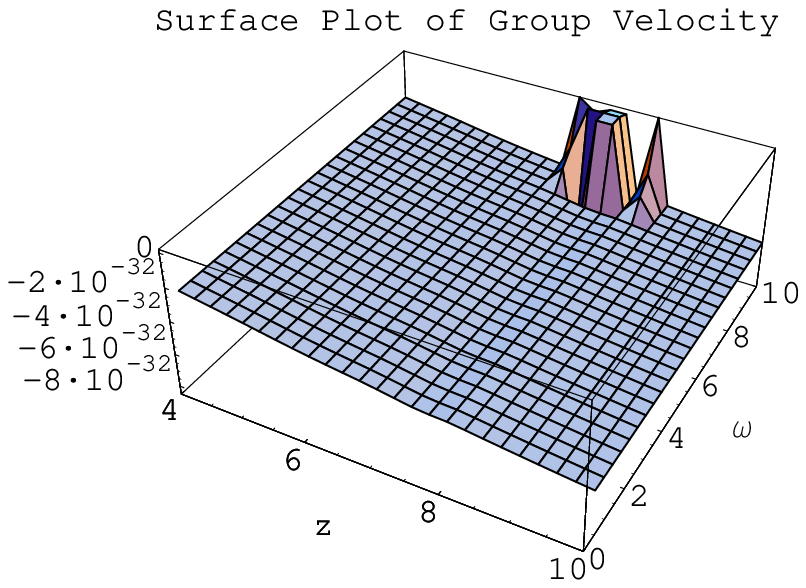,width=0.34\linewidth}\\
\end{tabular}
\caption{Waves exhibit normal as well as normal dispersion at random
points.}
\end{figure}

\begin{center}
Table V. Normal and anomalous dispersion regions
\begin{tabular}{|c|c|c|c|c|}
\hline \textbf{Fig.} & \textbf{ Normal dispersion} &
\textbf{Anomalous dispersion}
\\\hline \textbf{7}& $4\leq z\leq 5.1, 4\leq\omega\leq 6$ &
$4.9\leq z\leq 5.3, 0\leq\omega\leq 1.7$
\\\hline \textbf{8}& $8.1\leq z\leq 8.9, 1.5\leq\omega\leq 3.1$ &
$4.3\leq z\leq 5.1, 7\leq\omega\leq 8.3$
\\\hline \textbf{9}& $5.3\leq z\leq 5.4, 1.8\leq\omega\leq 1.9$ &
$5.1\leq z\leq 5.7, 7\leq\omega\leq 8$
\\\hline
& $4.1\leq z\leq 4.4, 1.5\leq\omega\leq 3$  &
$7\leq z\leq 7.9, .5\leq\omega\leq 1.1$     \\
\textbf{10}&   $7\leq z\leq 7.7, 4.2\leq\omega\leq 4.9$ &
\\\hline& $4\leq z\leq 4.4, 7\leq\omega\leq 7.7$  &
$8\leq z\leq 8.6, 9.2\leq\omega\leq 9.4$     \\
\textbf{12}&   & $4\leq z\leq 4.3, 0\leq\omega\leq .3$
\\\hline
\end{tabular}
\end{center}

\section{Summary}

Plasma modes of hot plasma around RN-dS horizon in
the presence of Veselago medium are investigated in this paper. We
rewrite perfect GRMHD equations in Veselago medium by implementing
ADM formalism. Plasma flow distracts due to gravitational effects of
black hole. Linear perturbation is applied on flow variables to
investigate the gravity effects of black hole. Moreover, it is
assumed that plasma flow is two dimensional, i.e., xz-plane.
Dimensionless quantities corresponding to flow variables are
introduced to build component form of linearly perturbed GRMHD
equations.

The technique of Fourier analysis is applied to form dispersion
relations for the rotating (non-magnetized and magnetized) plasma.
Non-magnetized hot plasma indicates that waves in Figure \textbf{1}
and \textbf{3} are directed towards black hole horizon and move
away from the horizon in Figure \textbf{2}. The dispersion is
anomalous throughout the region in Figures \textbf{1} and
\textbf{14} whereas it is normal in Figure \textbf{3}.

In
magnetized environment, waves are in the direction of event horizon
in Figures \textbf{4}-\textbf{6} and \textbf{8}-\textbf{9} while
they moves away from horizon in Figures \textbf{7} and \textbf{10}.
Normal as well as anomalous dispersion is observed in Figures
\textbf{7}-\textbf{10} while it is anomalous in Figure \textbf{5}
and normal in Figures \textbf{4} and \textbf{6}.

In conventional refraction, $n$ is always greater than one, however,
it is negative for Veselago medium. Phase velocity should be greater
than group velocity in unusual medium and so in all the Figures
refractive index is less than one. Index increases and decreases in
small regions, dispersion is normal when $\frac{dn}{d\omega}$ is
positive and anomalous otherwise. All the roots of dispersion
relation in non-magnetize and magnetize plasmas (isothermal and hot)
satisfy unusual properties of Veselago medium. Results reassert
presence of such unusual medium in RN-dS magnetosphere.

It is followed by previous work \cite{32} that inclusion of positive
cosmological constant in non-rotating black hole may reveal more
information about magnetosphere. Here we have inspected that
addition of charge and cosmological constant is beneficial to attain
normal dispersion at majority of points in the region, however, most
of the waves move towards event horizon. It may be concluded that
spacetime charge affects the electromagnetic field firmly that waves
are directed towards black hole horizon.

On comparison with recent manuscript \cite{34} on waves around RN horizon, it is observed
that addition of cosmological constant to charged black hole i.e., RN-dS spacetime provide
more interesting results in terms of energy extraction information around magnetosphere.
Hot plasma is more generalized form of plasma and it is reducible to isothermal plasma
when specific enthalpy is considered to be constant.

\renewcommand{\theequation}{A\arabic{equation}}

\section*{Appendix}

GRMHD equations are governed from basic MHD equations in account with Maxwells
electrodynamical equations.
$3+1$ GRMHD equations for RN-dS metric in a Veselago medium become \cite{32}.
\begin{eqnarray}{\setcounter{equation}{1}}
\label{4} &&\frac{\partial \textbf{B}}{\partial
t}=-\nabla\times(\alpha \textbf{V}\times \textbf{B}),
\end{eqnarray}
\begin{eqnarray}
\label{5}&&\nabla.\textbf{B}=0,\\
\label{6} &&\frac{\partial (\rho+p) }{\partial t}+(\rho+p)[\gamma^2
\textbf{V}. \frac{\partial \textbf{V}}{\partial t}+\gamma^2
V.(\alpha \textbf{V}.\nabla)
\textbf{V}+ \nabla.(\alpha\textbf{V})]=0,\\\label{7}
&&\left\{\left((\rho+p)\gamma^2+\frac{\textbf{B}^2}{4\pi}\right)\delta_{ij}
+(\rho+p)\gamma^4V_iV_j-\frac{1}{4\pi}B_iB_j\right\}
\left(\frac{1}{\alpha}\frac{\partial}{\partial
t}\right.\nonumber\\
&&\left.+\textbf{V}.\nabla\right)V^j+\gamma^2V_i(\textbf{V}.\nabla)(\rho+p)
-\left(\frac{\textbf{B}^2}{4\pi}\delta_{ij}-\frac{1}{4\pi}B_iB_j\right)V^j_{,k}V^k\nonumber\\
&&=-(\rho+p)\gamma^2a_i-p_{,i}+\frac{1}{4\pi}
(\textbf{V}\times\textbf{B})_i\nabla.(\textbf{V}\times\textbf{B})
-\frac{1}{8\pi\alpha^2}(\alpha\textbf{B})^2_{,i}\nonumber\\
&&+\frac{1}{4\pi\alpha}(\alpha B_i)_{,j}B^j-\frac{1}{4\pi\alpha}
[\textbf{B}\times\{\textbf{V}\times(\nabla\times(\alpha\textbf{V}\times\textbf{B}))\}]_i,\\
\label{8} &&-\frac{1}{\alpha}\frac
{\partial p}{\partial t}+(\rho+p)[(\frac{1}{\alpha}\frac{\partial}{\partial
t}+\textbf{V}.\nabla)\gamma^2+2\gamma^2(\textbf{V}.\textbf{a})
+\gamma^2(\nabla.\textbf{V})]\nonumber\\
&&
-\frac{1}{4\pi\alpha}(\textbf{V}\times\textbf{B}).[(\textbf{V}\times\frac
{\partial \textbf{B}}{\partial t})
+(\textbf{B}\times\frac{\partial
\textbf{B}}{\partial
t})-
(\nabla\times\alpha\textbf{B})]=0.
\end{eqnarray}
Component form of linearly perturbed GRMHD equation takes following form
\begin{eqnarray}\label{17}
&&\frac{1}{\alpha}\frac{\partial b_x}{\partial
t}-ub_{x,z}=(ub_x-vb_z-v_x+\lambda v_z)\nabla
\ln\alpha\nonumber\\
&&-(v_{x,z}-\lambda
v_{z,z}-\lambda'v_z+v'b_z+vb_{z,z}-u'b_x),\\\label{18}
&&\frac{1}{\alpha}\frac{\partial b_z}{\partial t}=0,\\\label{19}
&&b_{z,z}=0,
\\\label{20}
&&\rho\frac{1}{\alpha}\frac{\partial\tilde{\rho}}{\partial t}
+p\frac{1}{\alpha}\frac{\partial\tilde{p}}{\partial
t}+(\rho+p)\gamma^2v(\frac{1}{\alpha}\frac{\partial{v_x}}{\partial
t}+uv_{x,z})+(\rho+p)\gamma^2u\frac{1}{\alpha}\frac{\partial{v_z}}{\partial
t}\nonumber\\
&&+(\rho+p)(1+\gamma^2u^2)v_{z,z}=-\gamma^2u(\rho+p)[(1+2\gamma^2v^2)v'+2\gamma^2uvu']v_x \nonumber\\
&&+(\rho+p)[(1-2\gamma^2u^2)(1+\gamma^2u^2)\frac{u'}{u}-2\gamma^4u^2vv']v_z,
\\\label{21}
&&\left\{(\rho+p)\gamma^2(1+\gamma^2v^2)
+\frac{B^2}{4\pi}\right\}\frac{1}{\alpha}\frac{\partial
v_x}{\partial t}+\left\{(\rho+p)\gamma^4uv-\frac{\lambda B^2}{4\pi}\right\}\nonumber\\
&&\times\frac{1}{\alpha}\frac{\partial v_z}{\partial
t}+\left\{(\rho+p)\gamma^2(1+\gamma^2v^2)
+\frac{B^2}{4\pi}\right\}uv_{x,z}+\left\{(\rho+p)\gamma^4uv\right.\nonumber
\end{eqnarray}
\begin{eqnarray}
&&\left.-\frac{\lambda B^2}{4\pi}\right\}uv_{z,z}
-\frac{B^2}{4\pi}(1+u^2)b_{x,z}-\frac{B^2}{4\pi\alpha}\left\{\alpha'(1+u^2)+\alpha
uu'\right\}b_x\nonumber\\
&&+\gamma^2u(\rho\tilde{\rho}+p\tilde{p})\left\{(1+\gamma^2v^2)v'+\gamma^2uvu'\right\}
+\gamma^2uv(\rho'\tilde{\rho}+\rho\tilde{\rho}'\nonumber\\
&&+p'\tilde{p}+p\tilde{p}')+[(\rho+p)\gamma^4u
\left\{(1+4\gamma^2v^2)uu'+4vv'(1+\gamma^2v^2)\right\}\nonumber\\
&&+\frac{B^2u\alpha'}{4\pi\alpha}
+\gamma^2u(1+2\gamma^2v^2)(\rho'+p')]v_x
+[(\rho+p)\gamma^2\left\{(1+2\gamma^2u^2)\right.\nonumber\\
&&\left.(1+2\gamma^2v^2)v'-\gamma^2v^2v'
+2\gamma^2(1+2\gamma^2u^2)uvu'\right\}-\frac{B^2u}
{4\pi\alpha}(\lambda\alpha)'\nonumber\\
&&+\gamma^2v(1+2\gamma^2u^2)(\rho'+p')]v_z=0,
\\\label{22}
&&\left\{(\rho+p)\gamma^2(1+\gamma^2u^2)
+\frac{\xi^2B^2}{4\pi}\right\}\frac{1}{\alpha}\frac{\partial
v_z}{\partial t}+\left\{(\rho+p)\gamma^4uv -\frac{\xi B
^2}{4\pi}\right\}\nonumber\\
&&\times\frac{1}{\alpha}\frac{\partial v_x}{\partial t}
+\left\{(\rho+p)\gamma^2(1+\gamma^2u^2)+\frac{\xi^2B^2}{4\pi}\right\}
uv_{z,z}+\left\{(\rho+p)\gamma^4uv\right.\nonumber\\
&&\left.-\frac{\xi B^2}{4\pi}\right\}uv_{x,z}+\frac{\xi
B^2}{4\pi}(1+u^2)b_{x,z}+\frac{B^2}{4\pi\alpha}\left\{(\alpha\xi)'
-\alpha'\xi+u\xi(u\alpha'\right.\left.+u'\alpha)\right\}b_x\nonumber\\
&&+(\rho\tilde{\rho}+p\tilde{p})\gamma^2\left\{a_z
+uu'(1+\gamma^2u^2)+\gamma^2u^2vv'\right\}+(1+\gamma^2u^2)(p'\tilde{p}+p\tilde{p}')\nonumber\\
&&
+\gamma^2u^2(\rho'\tilde{\rho}+\rho\tilde{\rho}')+[(\rho+p)\gamma^4\{u^2v'(1+4\gamma^2v^2)+2v(a_z+uu'(1+\nonumber\\
&&2\gamma^2u^2))\}-\frac{\xi
B^2u\alpha'}{4\pi\alpha}+2\gamma^4u^2v(\rho'+p')]v_x+[(\rho+p)\gamma^2\{u'(1+\gamma^2u^2)(1+\nonumber\\
&&4\gamma^2u^2)+2u\gamma^2(a_z+(1+2\gamma^2u^2)vv')\} +\frac{\xi
B^2u}{4\pi\alpha}(\alpha\xi)'
+2\gamma^2u(1\nonumber\\
&&+\gamma^2u^2)(\rho'+p')]v_z=0,
\\\label{23}
&&\frac{1}{\alpha}\gamma^2\rho\frac{\partial \tilde{\rho}}{\partial
t}+\frac{1}{\alpha}\gamma^2p\frac{\partial \tilde{p}}{\partial
t}+\gamma^2(\rho'+p')v_z+u\gamma^2(\rho\tilde{\rho}_{,z}
+p\tilde{p}_{,z}+\rho'\tilde{\rho}+p'\tilde{p})\nonumber\\&&
-\frac{1}{\alpha}p\frac{\partial \tilde{p}}{\partial
t}+2\gamma^2u(\rho\tilde{\rho}+p\tilde{p})a_z
+\gamma^2u'(\rho\tilde{\rho}+p\tilde{p})+2(\rho+p)\gamma^4(uV'\nonumber\\
&&+2uva_z+u'v)v_x+2(\rho+p)\gamma^2(2\gamma^2uu'+a_z\gamma^4+2\gamma^2u^2a_z)v_z+\nonumber\\
&&2(\rho+p)\gamma^4uvv_{x,z}+(\rho+p)\gamma^2(1+2\gamma^2u^2)v_{z,z}
-\frac{B^2}{4\pi\alpha}[(v^2+u^2)\xi\frac{\partial b_x}{\partial
t}\nonumber\\
&&+(v^2+u^2)\frac{\partial b_z}{\partial t}-\xi
V(\xi v+u)\frac{\partial b_x}{\partial t}-u(\xi
v+u)\frac{\partial
b_z}{\partial t}]-\frac{B^2}{4\pi\alpha}\nonumber\end{eqnarray}
\begin{eqnarray}
&&\times[\xi(\xi v+u)v_{x,t}+(\xi v+u)v_{z,t}
-(\xi^{2}+1)vv_{x,t}-(\xi^{2}+1)uv_{z,t}]\nonumber\\
&&+\frac{B^2}
{4\pi}(\xi\xi'v_z-\xi'v_x-\xi'vb_z+\xi'ub_x-v
b_{x,z}+u\xi b_{x,z})=0.
\end{eqnarray}

\end{document}